\documentclass[a4paper,fleqn,usenatbib]{mnras}
\usepackage{color}
\usepackage{amsmath}
\usepackage{bm}
\usepackage{natbib}
\usepackage{hyperref}
\usepackage[capitalize]{cleveref}
\usepackage{graphicx}
\usepackage[dvipsnames]{xcolor}
\usepackage{soul}
\usepackage{ulem}

\newcommand{\update}[1]{#1}
\newcommand{\remove}[1]{}
\newcommand{\cmt}[1]{}

\newcommand{\newacronym}[3]{%
 \newcommand{#1}[1]{#3##1 (#2##1)%
 \renewcommand{#1}[1]{#2####1}}%
}

\newcommand{\renewacronym}[3]{%
 \renewcommand{#1}[1]{#3##1 (#2##1)%
 \renewcommand{#1}[1]{#2####1}}%
}

\newacronym{\BHNS}{BHNS}{black hole--neutron star}
\newacronym{\NSNS}{NSNS}{neutron star--neutron star}
\newacronym{\SPH}{SPH}{smoothed particle hydrodynamics}
\newacronym{\BH}{BH}{black hole}
\newacronym{\NS}{NS}{neutron star}
\newacronym{\sGRB}{sGRB}{short gamma ray burst}
\newacronym{\EOS}{EOS}{equation of state}
\newacronym{\NSE}{NSE}{nuclear statistical equilibrium}

\newcommand{\SkyNet}{\emph{SkyNet}}
\newcommand{\StarSmasher}{\emph{StarSmasher}}
\newcommand{\StarCrash}{\emph{StarCrash}}

\crefformat{footnote}{#2\footnotemark[#1]#3}
\crefname{equation}{Equation}{Equations}
\crefname{figure}{Figure}{Figures}

\newcommand\sref[1]{Section~\ref{#1}}

\newcommand{\cqg}{Class.~Quant.~Grav.}

\newcommand{\Caltech}{{TAPIR, Walter Burke Institute for Theoretical Physics, MC
350-17,
    California Institute of Technology, Pasadena, California 91125, USA}}

\newcommand{\WSU}{{Department of Physics \& Astronomy,
        Washington State University, Pullman, Washington 99164, USA}}

\newcommand{\LBL}{{Lawrence Berkeley National Laboratory,
1 Cyclotron Rd, Berkeley, California 94720, USA}}

\newcommand{\RIT}{Center for Computational Relativity and Gravitation and School
of Mathematical Sciences, Rochester Institute of Technology,
Rochester, \\ New York 14623, USA}
\newcommand{\Allegheny}{{Department of Physics, Allegheny College, Meadville,
Pennsylvania 16335, USA}}
\newcommand{\Guelph}{{Department of Physics, University of Guelph, Guelph,
Ontario N1G 2W1, Canada}}
\newcommand{\SciNet}{{SciNet HPC Consortium, University of Toronto, Toronto,
Ontario M5T 1W5, Canada}}

\title[Neutrinos and BHNS r-Process Nucleosynthesis]{The Influence of
Neutrinos on r-Process Nucleosynthesis in the Ejecta of
Black Hole--Neutron Star Mergers}

\author[L. F. Roberts et al.]{Luke F. Roberts$^{1\dag}$\thanks{email:lroberts@tapir.caltech.edu},
Jonas Lippuner$^{1}$,
Matthew D. Duez$^{2}$, 
Joshua A. Faber$^{3}$, 
\newauthor
Francois Foucart$^{4\dag}$,
James C. Lombardi Jr.$^{5}$,
Sandra Ning$^{1}$,
Christian D. Ott$^{1}$, 
\newauthor
and Marcelo Ponce$^{6,7}$
\\
$^{1}$ \Caltech \\
$^{2}$ \WSU \\
$^{3}$ \RIT \\
$^{4}$ \LBL \\
$^{5}$ \Allegheny \\
$^{6}$ \Guelph \\
$^{7}$ \SciNet \\
$^{\dag}$ NASA Einstein Fellow \\
}

\begin{document}

\maketitle

\begin{abstract}
During the merger of a black hole and a neutron star, 
baryonic mass can become unbound from the system. Because the ejected
material is extremely neutron-rich, the \mbox{r-process} rapidly synthesizes
heavy
nuclides as the material expands and cools. In this work, we map general
relativistic models of \BHNS{} mergers into a Newtonian \SPH{} code and follow
the evolution of the thermodynamics and morphology of the ejecta until the
outflows become homologous. We investigate how the subsequent evolution depends
on our mapping procedure and find that the results are robust. Using
thermodynamic histories from the \SPH{} particles, we then calculate the
expected nucleosynthesis in these outflows while varying the level of neutrino
irradiation coming from the postmerger accretion disk. We find that the ejected material
robustly produces r-process nucleosynthesis even for unrealistically high
neutrino luminosities, due to the rapid velocities of the outflow. Nonetheless,
we find that neutrinos can have an impact on the detailed pattern of the
r-process nucleosynthesis. Electron neutrinos are captured by neutrons to
produce
protons while neutron capture is occurring. The produced protons rapidly form
low mass seed nuclei for the \mbox{r-process}. These low mass seeds are
eventually
incorporated into the first \mbox{r-process} peak at $A \sim 78$\remove{,producing
mainly Ge
and Se}. We consider the mechanism of this process in detail and discuss if it
can impact galactic chemical evolution of the first peak r-process nuclei.
\end{abstract}

\renewacronym{\BHNS}{BHNS}{Black hole--neutron star}
\renewacronym{\SPH}{SPH}{smoothed particle hydrodynamics}


\section{Introduction}
\BHNS{} binary mergers are a likely candidate for Advanced LIGO and Advanced
VIRGO detections of gravitational waves \citep{aligo:15, avirgo:15}, they may be
responsible for \sGRB{s} \citep[e.g.][]{lee_rev:07}, and they may provide a
significant fraction of the r-process material found in our galaxy
\citep[e.g.][]{lattimer:76, korobkin:12, bauswein:14}. Within the next few
years, it is likely that Advanced LIGO will detect gravitational waves from
these systems and constrain the \BHNS{} merger rate. If electromagnetic
counterparts are detected, the merger-\sGRB{} connection may be confirmed and
production of the r-process nuclei may be observed {\it in situ}
\citep{metzger:12}.

The origin of the r-process nuclei has been a long standing question in nuclear
astrophysics \citep{burbidge:57}. Core-collapse supernovae are appealing as a
possible site because of galactic chemical evolution considerations
\citep[e.g.][]{qian:00,argast:04}, but there is significant difficulty finding
the requisite conditions for r-process nucleosynthesis in this environment
\citep[e.g.][]{arcones:13}. Conversely, it is
relatively easy to find conditions neutron-rich enough for r-process
nucleosynthesis in the material ejected from binary \NS{} and \BHNS{} mergers
\citep{freiburghaus:99}.  Due to the long delay time from binary formation to merger
and the large amount of material ejected per merger event, it is challenging to
get simple models of galactic chemical evolution, which invoke compact object mergers
for r-process production to agree with the observed distribution of r-process
elements in low metallicity halo stars \citep{qian:00,argast:04}.  Nevertheless,
recent works taking into account more complex models of galaxy formation get
reasonable agreement with the observed distribution of r-process elements
\citep{matteucci:14, shen:15, vandevoort:15, ishimaru:15} and it is possible to
get r-process enrichment at very low metallicity when different channels of
binary formation are considered \citep{ramirez-ruiz:15}.  Therefore, it is
plausible that compact object mergers could be \update{a significant
source of the galactic r-process nuclei.}

Recently, it has been recognized that weak interactions can significantly affect
the final composition of binary \NS{} outflows \citep{wanajo:14, goriely:15,
sekiguchi:15, foucart:15b, palenzuela:15, radice:16b}.  Likewise, the final state
and remnant product of binary \NS{} mergers has been shown to depend on several
properties of the system, e.g.\ important roles are played by the microphysical
nuclear \EOS{}, electromagnetic fields and neutrino effects \citep{neilsen:14,
palenzuela:15}.  In contrast to binary \NS{} mergers, the material ejected
during \BHNS{} mergers is unlikely to undergo significant numbers of weak
interactions.  Electron and positron captures are supressed relative to the
rates in the shock heated ejecta of binary \NS{} mergers due to the low entropy
present in the tidal ejecta. The high outflow speeds and low neutrino
luminosities encountered in these events---compared to binary \NS{}
mergers---also make it unlikely that neutrino interactions will drastically change the
number of neutrons present at the onset of r-process nucleosynthesis
\citep{foucart:14a,foucart:15}. Therefore, \update{the dynamical ejecta of} \BHNS{} mergers have been long
thought to be likely sites for production of heavy r-process nucleosynthesis
\citep{lattimer:76, lattimer:77, korobkin:12, bauswein:14b}. \remove{, although
calculations which do not include the effect of neutrinos show that only the
second and third r-process peaks are produced.}

Understanding how \BHNS{} mergers contribute to galactic chemical evolution
requires knowledge of the merger rate, predictions for the amount of mass
ejected per merger, the kinetic energy of the ejecta, and predictions of nuclei
synthesized in these outflows.  Although there are no observed \BHNS{} binaries,
theoretical predictions suggest that the rate of \BHNS{} binary mergers could be
up to a tenth of the rate of double neutron star binary mergers \citep{abadie:10,
bauswein:14b}. The amount of mass ejected during \BHNS{} mergers can depend
sensitively on the binary parameters, especially the \BH{} spin and mass
\citep{foucart:13, hotokezaka:13b, bauswein:14b, kyutoku:15}. More mass is
ejected as the \BH{} spin increases in the direction of the orbital angular
momentum \citep{foucart:14a}.  Increasing the spin decreases the radius of the
innermost stable orbit and decreases gravitational binding at the radius at
which the \NS{} is tidally disrupted. Increasing the \BH{} mass reduces the
amount of material remaining outside of the \BH{} after merger (for fixed \NS{}
properties), since the tidal radius scales as \smash{$(M_{\rm BH}/M_{\rm
NS})^{1/3}R_{\rm NS}$} while the innermost stable orbit of the \BH{} scales as
$M_{\rm BH}$ for fixed \BH{} spin. The fraction of the mass outside the horizon
which is unbound, however, also increases with the \BH{} mass, making the
relation between \BH{} mass and unbound mass nontrivial \citep{kyutoku:15}.  

Because the mass and spin distributions of stellar mass \BH{s} and the
expected number of \BHNS{} system in our galaxy are
not well known \citep[e.g.][]{ligo:10}, it is difficult to estimate the
contribution of these events to the r-process material found in the galaxy
\citep{bauswein:14b}. Nonetheless, it is timely to investigate the detailed
composition of the ejecta because the merger rate is likely to soon be
constrained by Advanced LIGO \citep{aligo}. Additionally, there are some hints
that the infrared excess associated with GRB130603B \citep{tanvir:13, berger:13}
is consistent with that event being powered by the radioactive decay of
r-process products in the ejecta of a \BHNS{} merger \citep{hotokezaka:13b}.
A similar excess has recently been observed in the afterglow of 
GRB060614 \citep{yang:15, jin:15}.

In this work, we investigate the long term hydrodynamics of the \BHNS{} ejecta
and the nucleosynthesis that occurs therein.
For the first time, we focus on how neutrinos might affect the
detailed nucleosynthesis patterns that are produced. Even for unrealistically
large neutrino luminosities, we find that the distribution of the pre-neutron capture
electron fraction is not significantly altered and the second and third
r-process peaks are robustly
produced in almost all of the material. This is in contrast to the dynamical
ejecta of binary \NS{} mergers, where weak processing may prevent an r-process
from occurring in a significant amount of the material \citep{wanajo:14,
goriely:15}. Of course, the \BHNS{} result is expected because the outflows
happen
relatively early before the remnant disk can start to emit neutrinos, there is
no hypermassive \NS{} contributing to the neutrino flux, and the tidal ejecta
possesses a very high velocity. More interestingly, we find that electron
neutrino captures by neutrons can provide seed
nuclei for a low mass r-process that produces material in the first r-process peak at
$A\sim78$.  Nonetheless, in our models, the ratio of the first peak to the
second peak is sub-solar with and without the inclusion of neutrino captures.
When comparing to the yields of low metallicity halo stars with
sub-solar Ge abundances \citep{roederer:14}, we find that this first peak
production can bring our models closer to agreement with the observed abundances
of Ge, As, and Se, although the abundances are still somewhat low.

This paper is organized as follows: in \sref{sec:methods},
we present the \BHNS{} systems we have simulated,
explain how the ejected material is mapped into our \SPH{} code, and describe
our nuclear reaction network. Then, in \sref{sec:ye}, we discuss the
effect of weak interactions on the electron fraction distribution in the ejecta.
In \sref{sec:first_peak}, we present the integrated nucleosynthesis from
our models and discuss neutrino induced production of the first r-process peak.
In \sref{sec:GCE}, we discuss uncertainties in the results from our
nucleosynthesis calculations and their possible implications for galactic
chemical
evolution and for abundance observations in low metallicity halo stars.

\section{Methods}
\label{sec:methods}

\subsection{Relativistic Merger Simulations and Binary Systems}
\label{sec:binaries}
The \BHNS{} merger simulations used in this work have been described in detail
in
our previous papers \citep{deaton:13,foucart:14a}.
Here we review the major features and error estimates of the merger simulations,
referring readers to \cite{foucart:14a} for details. The fully relativistic
Einstein-hydrodynamics system is evolved with the Spectral Einstein Code (SpEC)
\citep{SpEC}.  Neutrino cooling and lepton number evolution are incorporated
through a neutrino leakage scheme \citep{deaton:13}.

To model the \NS{}, we employ the Lattimer-Swesty \EOS{} \citep{lseos:91} with
an incompressibility
$K_0=220\,{\rm MeV}$ and a symmetry energy $S_\nu=29.3\,{\rm MeV}$ (hereafter
LS220), using the table available at \url{http://www.stellarcollapse.org} and
described in \cite{oconnor:10}. This \EOS{}
yields a neutron star radius that lies within the allowed range of radii, as
determined by \cite{hebeler:13} from nuclear theory constraints
and the existence of neutron stars of mass $\sim
2M_\odot$ \citep{demorest:10b,antoniadis:13}. For LS220, a 1.2
(1.4)$M_{\odot}$ neutron star has a radius $R_{\rm NS}$ of 12.8 (12.7)\,km and a
compactness \smash{$C=GM_{\rm NS}/(R_{\rm NS}c^2)$} of 0.139 (0.163).

During the SpEC simulations, the dynamical ejecta is tracked for only about
5\,ms before it exits
the computational grid. However, during this time, the specific energy ($u_t$)
of fluid elements becomes nearly constant, so it is often possible to
confidently identify unbound material.
Convergence of our SpEC simulations was observed to be faster than
second-order.
Assuming second order convergence gives a conservative relative error of up to
60\% in the mass and kinetic energy of ejected material. Even if the true error were
this large, which is unlikely, it would not affect the results of the
present investigation. As we will see, variations of ejecta properties
between different binary systems, which are of similar magnitude, have
negligible effect on the final nuclear abundances.

In the simulations of \cite{deaton:13} and \cite{foucart:14a}, we considered
\BHNS{} binary systems with multiple masses and spins.  The
\BH{} mass \smash{$M_{\rm BH}$} was taken to be \smash{5.6$M_{\odot}$},
\smash{7$M_{\odot}$}, or \smash{10$M_{\odot}$}, covering most of the estimated
mass distribution for stellar mass black holes \citep{ozel:10bh, farr:11}.  The
neutron star gravitational mass \smash{$M_{\rm NS}$} was taken to be
\smash{1.2$M_{\odot}$} or \smash{1.4$M_{\odot}$}, which is typical for \NS{s}
 \citep{kiziltan:13}.
For these masses, ejecta is produced only for at least moderately high \BH{}
spins, meaning that for most cases the Kerr spin parameter must be
\smash{$\chi_{\rm BH} > 0.7$} \citep{foucart:12dm}.

For this study, we use the ejecta from three systems.
The first, called
``M12-7-S9'', with parameters $M_{\rm NS}=1.2M_{\odot}$, $M_{\rm
BH}=7M_{\odot}$, $\chi_{\rm BH}=0.9$, produces a very large ejecta mass of
0.16$M_{\odot}$. The second, ``M14-7-S8'', with $M_{\rm NS}=1.4M_{\odot}$,
$M_{\rm BH}=7M_{\odot}$, $\chi_{\rm BH}=0.8$, has ejecta mass 0.06$M_{\odot}$,
one of our lower ejecta mass cases. The third case, ``M14-5-S9'', has parameters
$M_{\rm NS}=1.4M_{\odot}$, $M_{\rm BH}=5.6M_{\odot}$, $\chi_{\rm BH}=0.9$ and
ejects a mass of 0.08$4M_{\odot}$.

\subsection{SPH evolution of ejecta}

After $\sim$5\,ms, the ejecta has detached from the merger remnant and
is moving for the most part ballistically. However, the outflow
is not yet homologous. Also, it is possible that pressure forces
will subsequently become important again because of recombination
heating or collision of streams of matter (although this turns out
not to be the case). Therefore, we continue the hydrodynamic
evolution of the outflow using an \SPH{}
code, \StarSmasher\ \citep{Gaburov:2009kg,Ponce:2011kv}. The \SPH{} code is
Newtonian, but since the flow
is only mildly relativistic ($v/c\approx 0.2$), and from the beginning
somewhat far from the black hole ($>10M_{\rm BH}$), this is probably
adequate for our purposes. (See check on this below.)

The \StarSmasher\ code  is the successor
to the earlier \StarCrash\ code \citep{Lombardi:2005nm}.
It represents fluids in the standard \SPH{} way, using a finite number
of fluid elements or ``particles.''  In its current implementation,
the particles may have different masses \citep{Gaburov:2009kg}, which
simplifies the construction of initial data from finite volume
representations.
\StarSmasher\ uses variable smoothing lengths to maximize resolution,
using a formalism derived consistently from a particle-based Lagrangian
to ensure proper energy and entropy evolution \citep{Lombardi:2005nm,
Springel:2001qb,Monaghan:2002ru}.

Stable shock evolution is achieved using artificial viscosity with a
Balsara switch \citep{Balsara1995} to suppress artificial viscosity
in shear layers; fortunately, accurate shock evolution is not
important for our application.  Self-gravity forces are neglected,
so the gravitational force is simply a function of position given
by the black hole potential and it is implemented in the
Newtonian and Paczy\'{n}ski-Wiita approximations \citep{Ponce:2011kv,
Paczynski:1979rz}. In order to
avoid small time steps due to rapid motion,
particles are removed if they come too close to the \BH{}.  These particles
would eventually fall into the \BH{} anyway, so this procedure does not affect
the ejecta properties.

As initial data to the \SPH{} simulation, hydrodynamic data from a
snapshot of the SpEC merger simulation (taken after tidal disruption
but before the tidal tail hits the outer boundary) are output on
a uniform Cartesian mesh. \StarSmasher\ reads these data, reflects them
to add the lower hemisphere not evolved by SpEC, interpolates to an
hexagonal close-packed lattice, and assigns a particle of
appropriate density to each nonvacuum lattice point.
The evolution is then continued in \StarSmasher\ using
the LS220 \EOS{} with no neutrino effects. The electron fraction
$Y_e$ of each particle is constant during the \SPH{} evolution, and no neutrino
cooling or absorption is considered. If a particle falls below the LS220
density or temperature
table range, the entropy $S$ is henceforth set to be constant, and a
$S\propto \rho T^3$, $P\propto \rho^{4/3}$ extrapolation of the \EOS{} is used.
This only happens when pressure is negligible, and the entropy evolution in
\StarSmasher\ is not used in our post-processing nucleosynthesis calculations
(see below).

Although relativistic effects are not expected to be important, the
translation from relativistic to Newtonian physics must account for
two subtleties. First, the late-time behavior of an ejecta fluid element
is most sensitive to its energy, especially whether it is bound or unbound,
so it is important that this be appropriately translated. We therefore
rescale the velocity vector so that the specific kinetic plus potential
energy of each particle in the Newtonian framework is equal to its
relativistic specific energy $-u_t-1$ in the SpEC simulation. Second,
there is no a priori guarantee that the coordinate system in which
the numerical relativity simulation evolves will be close to any
known coordinates. Fortunately, SpEC's ``damped harmonic'' coordinates
lead the spacetime to settle nearly in harmonic coordinates, so we
transform in \StarSmasher\ to Schwarzschild coordinates (ignoring the
\BH{'s} spin, whose effects will not be important far from the hole),
with a simple radial transformation $r\rightarrow r+M_{\rm BH}$. In the
region of interest, the numerically evolved spacetime is nearly Minkowski,
and the deviation from Minkowski is mostly Schwarzschild and so can be
adequately modeled by a Paczy\'{n}ski-Wiita potential. Lastly, one
must distinguish between the rest frame baryon density used in the \EOS{} and
the mass integrand density used to assign the mass of the \SPH{} particle, which
is the rest frame baryon density times a Lorentz factor and a metric
determinant factor. Because \SPH{} particle mass is a constant, the mass
integrand density only needs to be calculated and integrated over at the initial
time.

A straightforward evolution of the fluid equations produces a generally
realistic evolution but with some clearly unphysical artifacts. Namely,
matter on the upper and lower surfaces of the ejected tidal tail blow away from the
equator, something unexpected given the overall weakness of pressure forces
and not indicated in the SpEC evolution. This vertical expansion has no
influence on the energy distribution or nucleosynthesis results, but it
does affect the shape of the outflow. Convergence tests show that it is
not a transient caused by an insufficient number of particles, so
it is likely an artifact of the transition to Newtonian physics. It can be removed
by reducing pressure forces near the black hole. In our simulations labeled
``P1'', we turn off pressure forces within 10$M_{\rm BH}$ of the black hole,
while within 100$M_{\rm BH}$, pressure is reduced by a factor varying linearly
with distance between zero (at 10$M_{\rm BH}$) and one (at 100$M_{\rm BH}$).
Within this range, the specific entropy is held constant, because otherwise
the pressure reduction would keep the fluid from adiabatically cooling.
Simulations labeled ``P2'' have full pressure forces everywhere.

\begin{figure}
\includegraphics[width=\linewidth]{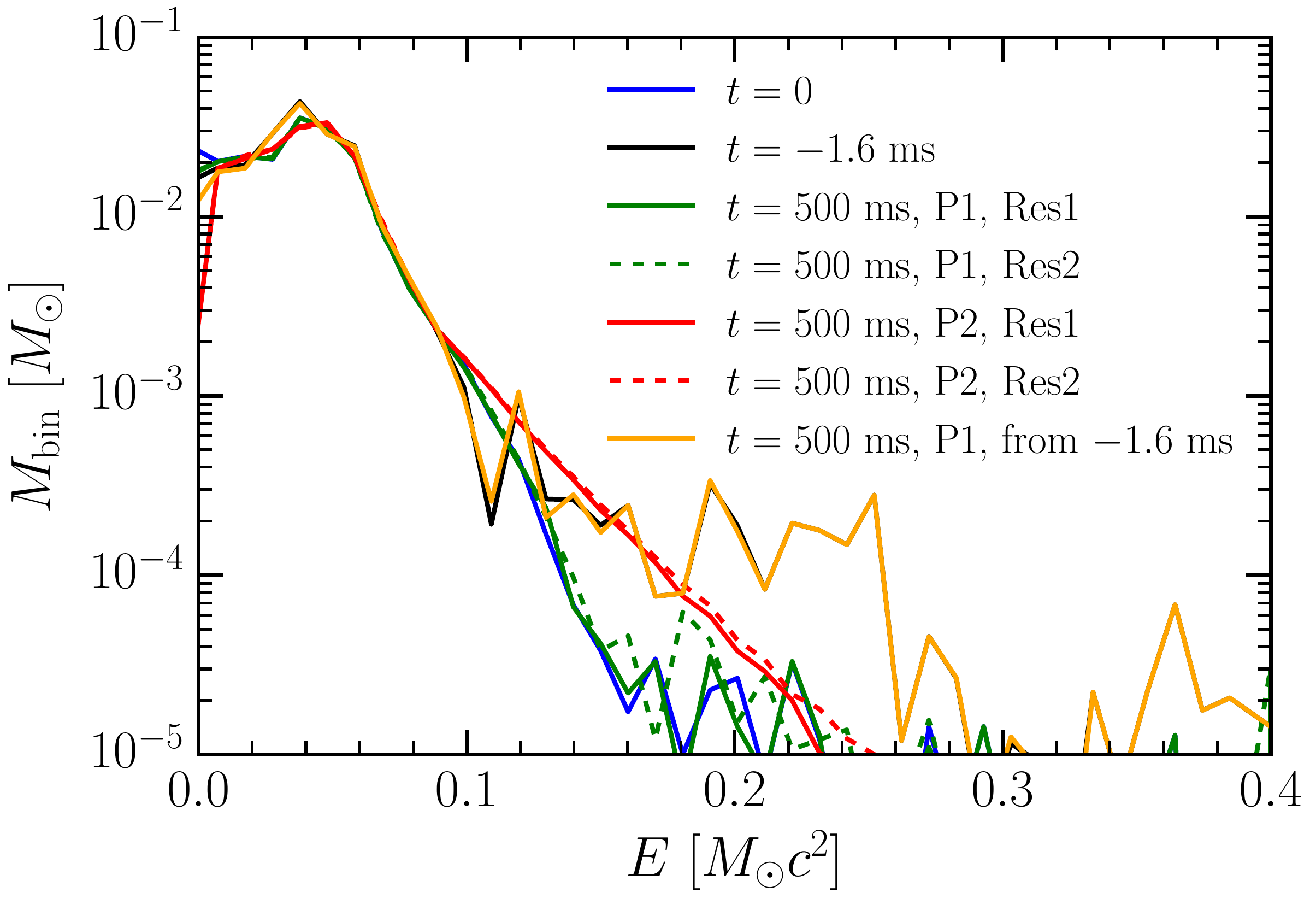}
\caption{The distribution of specific kinetic plus potential energy in the
unbound post-merger matter, shown for system M12-7-S9 at a time shortly after
the disruption of the \NS{} ($t=0$) and 500ms later, long after the distribution
has settled. For each energy bin, we integrate the density of all particles with
energy inside that bin, giving a Newtonian mass for each energy bin.  We show 2
resolutions, ``Res1'' and ``Res2'', corresponding to around 79,000 and 175,000
particles, respectively.  We evolve using two methods: ``P1'' turns off pressure
forces and imposes adiabatic internal energy evolution within a radius of about
$100M_{\rm BH}$. ``P2'' includes pressure forces everywhere but removes bound
particles after 10~ms. Another \SPH{} run begun 1.6\,ms earlier in the merger
has nearly stationary energy distribution if evolved with P1. A simulation using
P2 with a Paczy\'{n}ski-Wiita potential gives results almost identical to P2
with the standard Newtonian point potential.}
\label{fig:energy}
\end{figure}

We check that our evolved results are insensitive to the time at which
we transition from SpEC/relativistic to \StarSmasher/Newtonian physics by
starting from two different snapshots 1.6\,ms apart and finding negligible
variation in the evolved energy histograms, which are shown in 
\cref{fig:energy}. In fact, even the initial
energy histograms are not very different, so after the first $\approx 2$\,ms
the influence of pressure on the kinematics of the ejecta is negligible. The
codes' main function is to provide the density evolution as particles follow
their ballistic trajectories.

Our interest is only in unbound matter. The bound matter for the most
part orbits the black hole in an accretion disk or is ``eaten'' when
it comes within the prescribed distance from the central point mass.
Both because of the exclusion of full relativity and the lack of a
transport process to drive accretion, the disk evolution cannot be
regarded as believable. We find that, for P2 evolutions, if we allow
the disk to evolve for long periods of time, some fraction of the mass
becomes weakly unbound. This is not perhaps incorrect given the physics
included,
but it cannot be regarded as physical, so we remove this contamination
by eliminating bound particles after 10\,ms of \SPH{} evolution. For P1
evolutions,
this removal is not necessary.

\begin{figure}
\includegraphics[width=\linewidth]{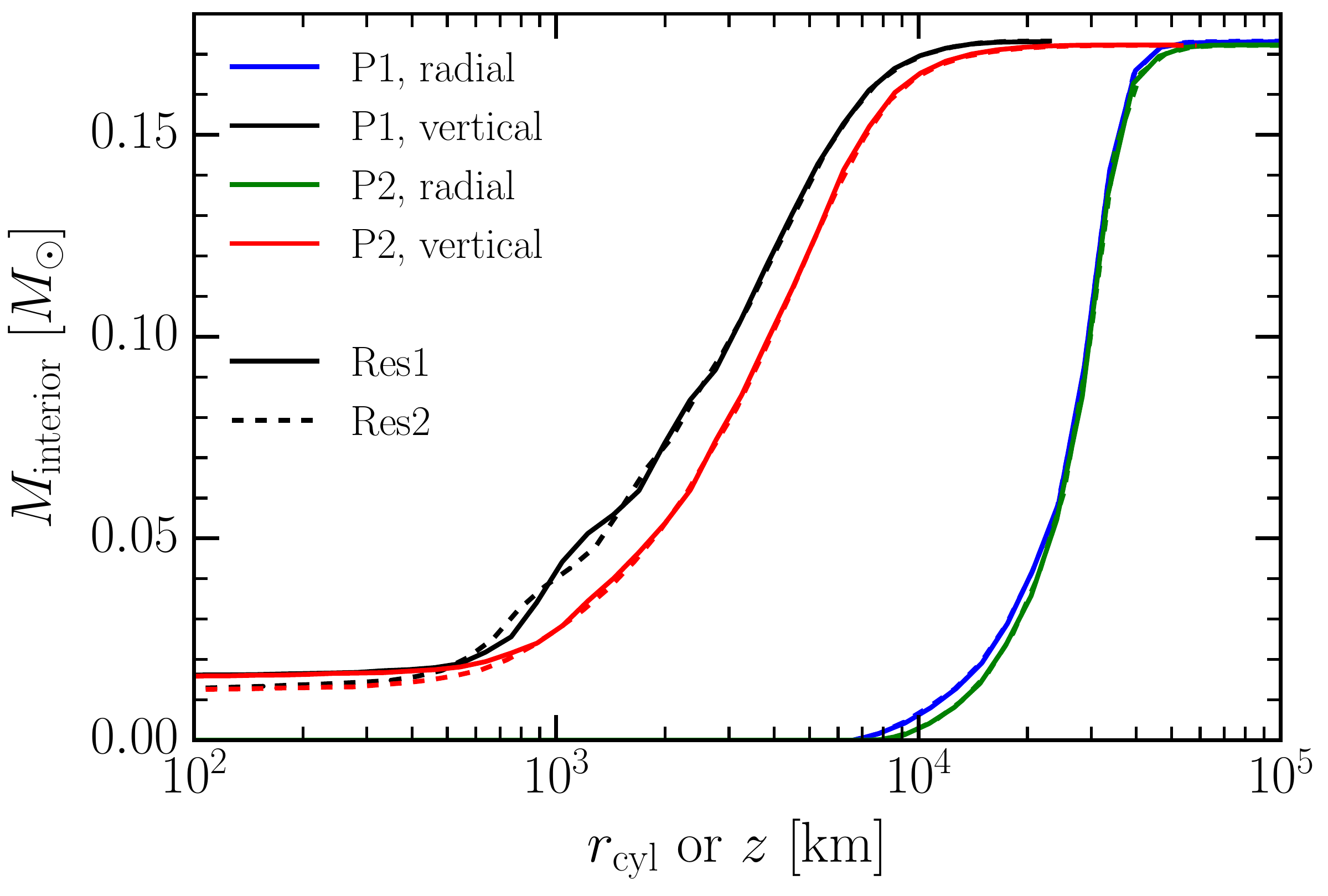}
\caption{The mass (computed by density integral) interior to a given
 cylindrical radius $r_{\rm cyl}$ or vertical height $|z|$ for binary
 ejecta M12-7-S9. Profiles
 are computed at a time 500~ms after merger, by which point the
 ejecta profile has settled and will thereafter spread nearly
 homologously. The vertical
 interior mass appears to asymptote to a nonzero value on the left,
 indicating that a significant number of particles remain near the
 equator. We show results for two resolutions with two ways of
 handling pressure forces. Simulations with pressure forces completely
 turned off give profiles nearly the same as P1 profiles.}
\label{fig:Mint}
\end{figure}

\begin{figure}
\includegraphics[width=\columnwidth]{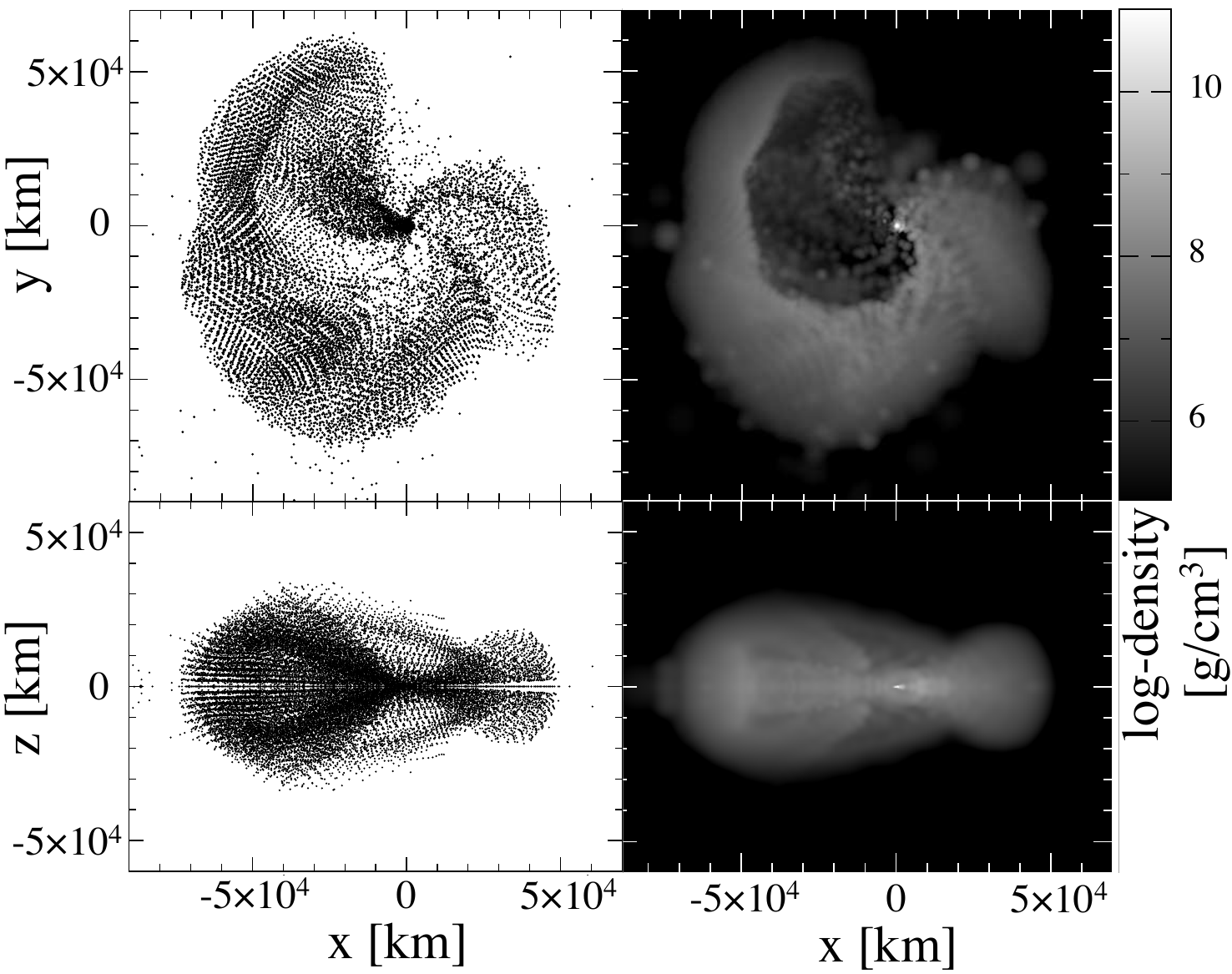}
\caption{Outflow's profile at a representative snapshot from the SPH evolution
 of M12-7-S9. 
Upper/bottom-left panels: $xy$/$xz$-projection of SPH-particles;
upper/bottom-right panels: density (in log-scale [g cm$^{-3}$]) views in the
$xy$/$xz$-planes respectively.
}
\label{fig:sph-splash} \end{figure}

Our standard evolutions use roughly 75,000 particles.
The mass profile of the ejecta M12-7-S9
is shown in \cref{fig:Mint}.  \Cref{fig:sph-splash} shows snapshots
of the \SPH{} particles and fluid density after $\approx0.5\ \text{s}$ of
starting the \SPH{} evolution.  

\subsection{Nuclear reaction network and weak interactions}

To calculate the composition of a Lagrangian fluid element in the ejecta, we
require the evolution of its density as a function of time as well as its
initial composition and entropy.  To allow evolution at very late times, we
extrapolate the density histories of the particles taken from the \SPH{}
simulation assuming homology, $\rho \propto t^{-3}$, \update{which accurately
describes the long term evolution of the flow}. In
addition to the density, we extract the entropy and electron fraction along
these trajectories. The extracted electron fraction is constant due to the
neglect of weak reactions during the \SPH{} evolution. The LS220 \EOS{} is only
valid for baryon densities and temperatures above \smash{$\rho = 10^{8}\,
\textrm{g cm}^{-3}$} and $T \approx 1\, \textrm{GK}$, which does not include the
entire region in the temperature density plane in which neutron capture occurs.
Since corrections to the \EOS{} due to nuclear interaction become negligible
below \smash{$\rho \sim 10^{12}\, \textrm{g cm}^{-3}$}, we switch from LS220 to
a multi-species non-degenerate ideal gas \EOS{} consistent with the nuclei
included in our network along with the electron \EOS{} of \cite{timmes:99} for
our network evolutions. When keeping the entropy fixed and assuming an initial
\NSE{} composition---with the modified Helmholtz \EOS{} described in
\cite{lippuner:15}---we find temperature differences less than a few percent
between the two \EOS{} in the region where they overlap. \update{Because the
single nucleus approximation of LS220 predicts different nuclei than a full NSE
calculation, there is a mismatch between the total internal energies of $\sim 0.1 \,
\textrm{MeV/baryon}$ when switching between the two EOSs. This level of error is
unlikely to signicantly impact the nucleosynthesis calculations because the
total energy released per baryon during the nucleosynthesis is of order ten MeV.}

Once the density evolution of the Lagrangian particles has been extracted and
extrapolated, we evolve the composition of the particles using the nuclear
reaction network code \SkyNet\ \citep{lippuner:15} and the network described
therein. \remove{The nuclear network
employed includes 7843 different isotopes extending from neutron to
$^{337}_{112}$Cn. We employ both the forward reaction rates and the nuclear
data tabulated in REACLIB 
. Inverse reaction rates are
calculated assuming detailed balance. In addition to the REACLIB reactions,
neutron induced fission rates from 
and spontaneous
fission rates calculated from the approximation of 
using the
spontaneous fission barriers of 
are included. Symmetric
fission fragments are assumed.}
The entropy generated via nuclear transmutations is self-consistently included
in the evolution, similarly to \cite{freiburghaus:99}. At 3~ms after
merger---the time at which the SpEC simulations are mapped to
\StarSmasher---the
particles are typically at temperatures over 10~GK and \update{densities are below
$10^{12} \, \textrm{g cm}^{-3}$}. At these temperatures
\update{and
densities}, \NSE{}
holds, but weak interactions are generally far from equilibrium \update{over the
short timescales encountered during the merger}. To follow
changes in the electron fraction at high temperature, \SkyNet\ includes an
\NSE{}
evolution mode where strong interactions are assumed to be in equilibrium and
only weak interactions are tracked. This mode is used until the temperature
drops below 7~GK, at which point the full nuclear reaction network is evolved.
Because inverse strong reactions are calculated via detailed balance, the
transition
between the two \SkyNet\ evolution modes is smooth.

To track the potential importance of neutrino irradiation of the ejecta,
electron
neutrino capture, electron antineutrino capture, electron capture, and positron
capture by free nucleons are included in both evolution modes.
The neutrino capture rates are given by
\begin{align}
\lambda_\nu &= \frac{G_F^2 (1 + 3g_A^2)}{2 \pi^2 \hbar^7 c^6} \nonumber\\
&\phantom{=}\times\int_{\tilde Q}^{\infty} d\epsilon_e p_e \epsilon_e
(\epsilon_e - Q)^2 \bar f_\nu(\epsilon_e - Q) (1 - f_e(\epsilon_e)),
\end{align}
where $f_e$ is the electron distribution function, $G_F$ is the Fermi coupling
constant, $g_A$ is the weak axial vector coupling constant, $\epsilon_e$ is the
electron energy, $p_e$ is the electron momentum, $Q$ is the energy transfer from
the nucleons to the final state electron, and $\bar f_\nu$ is the angle-averaged
neutrino distribution function. The $Q$-value is defined in the direction of
electron or positron capture and $\tilde Q = \max(Q,m_e c^2)$.  The electron and
positron capture rates, $\lambda_{e^+}$ and $\lambda_{e^-}$, are calculated from
similar expressions with the distribution functions interchanged. This
expression assumes there is no momentum transfer to the nucleons and neglects
weak magnetism corrections. Although these corrections are potentially
significant in the case of neutrino driven winds \citep{horowitz:02}, they are
unlikely to significantly affect the evolution of the electron fraction in
\BHNS{} merger ejecta. The $\alpha$-effect locks free protons in heavy nuclei
and thereby prevents significant competition from electron antineutrino capture
\citep{fuller:95}.  We assume that the neutrino distribution has a Fermi-Dirac shape
in energy space and neutrinos of all energies are emitted from a single
spherical surface, which results in the distribution function
\begin{equation}
f_\nu(\epsilon, \mu, r) = 
\frac{\theta(\mu - \mu_0(r))}{\exp(\epsilon/T_\nu) + 1},
\end{equation}
where $\mu$ is the cosine of the angle of neutrino propagation relative to the
radial direction, \smash{$\mu_0 = \sqrt{1-(r_\nu/r)^2}$}, $T_\nu$ defines the
neutrino spectral temperature, $\theta$ is the Heaviside step function,
$\epsilon$ is the neutrino energy, and $r_\nu$ is the radius of neutrino
emission.  Inside of $r_\nu$, $\mu_0$ is assumed to smoothly approach negative
one over a tenth of $r_\nu$. The value of $r_\nu$ can be fixed by choosing a
neutrino luminosity, $L_\nu$, and spectral temperature. This model is crude,
considering the disk like geometry of the neutrino emitting region, but it is
sufficient for this study given that we are parameterizing the properties of the
neutrino field anyway.  In the following sections, we consider models with fixed
electron neutrino luminosities of $L_{\nu_e} = \{0, 0.2, 1, 5, 25\} \times
10^{52} \, \textrm{erg s}^{-1}$. The electron antineutrino luminosity is always
fixed to be $L_{\bar \nu_e}=1.5 L_{\nu_e}$, but our results are insensitive to
this choice due to the $\alpha$-effect.  These values are in the range found
in the simulations of \cite{foucart:15b} and the difference between the values
accounts for re-leptonization of the disk.  Since only charged current
interactions are included in the nuclear network, the properties of the heavy
flavored neutrino fields do not affect our results.  We employ constant
luminosities to reduce the number of parameters affecting our nucleosynthesis
calculations.

\remove{Where available, beta-decay and electron capture rates from
and
are used. For nuclei for which these rates are not available,
the effects of electron blocking and positron capture are approximately included
by assuming that the entire beta-decay strength is provided by a ground state to
ground state transition as described in 
. The matrix element
is chosen such that the beta-decay rate in vacuum is equal to the REACLIB
beta-decay rate. This procedure assumes a maximal Q-value and therefore
provides a lower limit on the importance of medium dependent effects.}

We perform nucleosynthesis postprocessing for all of the ejected \SPH{}
trajectories.  \remove{The network integration begins at three milliseconds after
merger. The initial conditions are specified by the density and electron
fraction at which this temperature is reached and by \NSE{}. The nuclear
abundances are then evolved in time along with the entropy of the fluid element,
which is self-consistently evolved due to nuclear transmutation.} The nuclear
evolution is followed until $10^{13}$~s after the merger, which allows for the
decay of all but a handful of long lived unstable isotopes.

\section{Results and Discussion}
\label{sec:results}

\subsection{The Electron Fraction of the Ejecta}
\label{sec:ye}

\begin{figure}
\includegraphics[width=\linewidth]{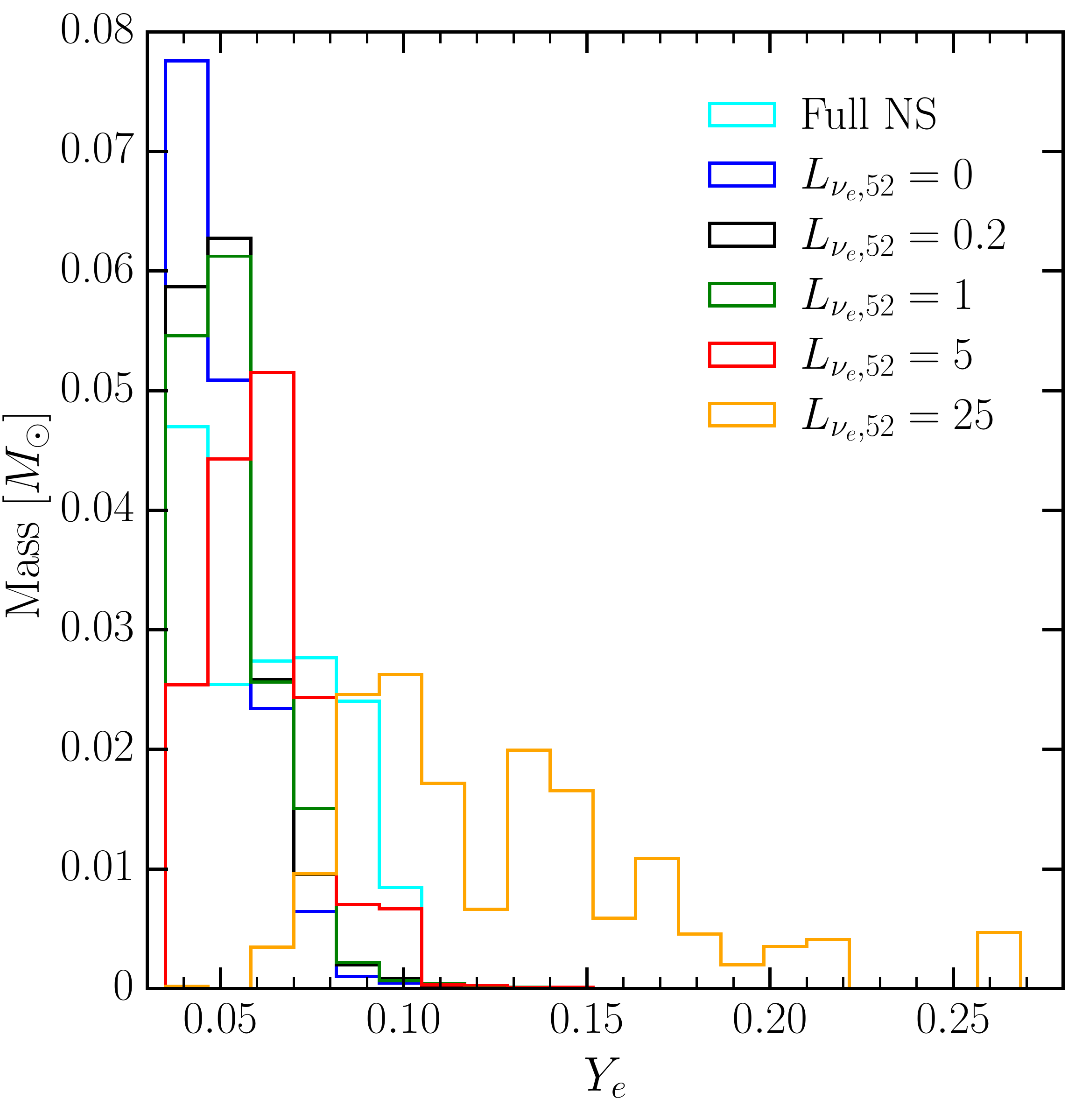}
\caption{Mass weighted histogram of the electron fraction in the ejecta from
model M12-7-S9 assuming fixed electron neutrino luminosities of $\{0, 0.2, 1, 5,
25\} \times 10^{52}\text{\ erg s}^{-1}$.  For comparison, we also show the
electron fraction histogram in a $1.2 \, M_\odot$ LS neutron star (cyan line).}
\label{fig:ye_distribution} \end{figure}

\begin{figure}
\includegraphics[width=\linewidth]{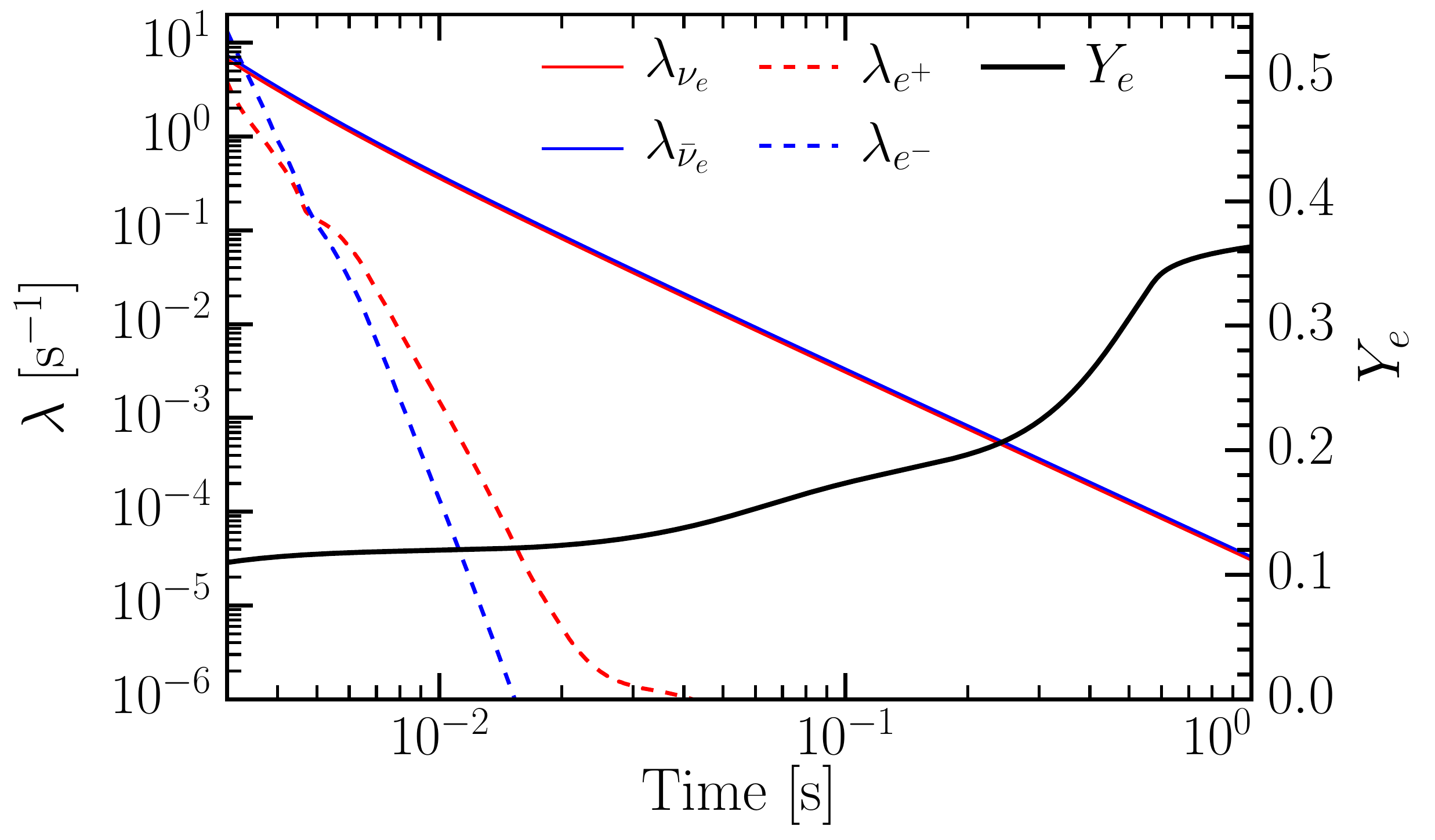}
\caption{Evolution of the electron fraction and weak rates as a function of time for a characteristic fluid
element. The electron neutrino luminosity is assumed to be $10^{53} \,
\textrm{erg s}^{-1}$. Because of the relatively low entropy of the \BHNS{}
ejecta and because of the low initial density of our calculations, neutrino
interaction rates dominate the electron and positron capture rates but neither
have a large impact on the electron fraction of the outflow. The increase in
$Y_e$ seen after around 100 ms is due to beta-decay during the r-process.}
\label{fig:weak_rates}
\end{figure}

The electron fraction of the material ejected during the \BHNS{} merger is the
most important parameter in determining the nucleosynthesis that occurs within the
outflow \citep[e.g.,][]{lippuner:15}. Given the short dynamical timescales and
the lack of a hypermassive central \NS{} after the merger, it has often been
assumed that the electron fraction of the dynamical ejecta from \BHNS{} mergers
is set solely by the initial beta-equilibrium electron fraction of the \NS{}
from which the material was ejected \citep{just:15}. If there are not a
substantial number of weak interactions during and after the merger, the
electron fraction will be low enough that an r-process involving a significant
number of fission cycles will occur: the outer layers of a \NS{} have $Y_e<0.1$
and the critical value for producing r-process material at low entropy is $Y_e
\approx 0.25$ \citep[e.g.][]{kasen:15, lippuner:15}. Neutrinos can impact the
electron fraction of the ejecta of binary \NS{} mergers \citep{wanajo:14,
goriely:15, foucart:15b, palenzuela:15, radice:16b}.  In binary \NS{}
mergers, a large fraction of the prompt ejecta comes from the shock heated
material in the interaction region of the two \NS{s} \citep{palenzuela:15}. The
increased temperatures and the large neutrino fluences near this material
increases $Y_e$ significantly and can sometimes drastically alter the character
of nucleosynthesis in the outflow. In the \BHNS{} case, there is no interaction
region during the tidal disruption of the \NS{}, and matter ejection when the
tidal stream self-intersects is very subdominant~\citep{foucart:15}.  The case
M14-5-S9 has the most massive ejection from the tidal stream
collision~\citep{deaton:13} of these \BHNS{}, but even for this case the imprint
of this secondary ejecta source on the overall outflow composition is small.
Therefore, the ejected material has a lower average entropy and electron
fraction than \NSNS{}  merger ejecta and there is no significant neutrino
emission until a disk has formed around the \BH{}.  Here, we consider the extent
to which neutrino interactions can alter the distribution of $Y_e$ just before
r-process nucleosynthesis begins in the ejecta.  

We estimate the effect of neutrino captures on the \BHNS{} outflows by
considering the maximum disk neutrino luminosities found by \cite{foucart:14a}.
The neutrino luminosity coming from the disk in both electron neutrinos and
antineutrinos is around $10^{53} \, \textrm{erg s}^{-1}$. Although the simulations
of \cite{foucart:14a} used a gray leakage approximation, we can get some
estimate of the average neutrino energies from the temperature of the emission
region which was around $5 \, \textrm{MeV}$, which suggests average neutrino
energies around $\epsilon_\nu \approx 3.15 T \sim 15 \, \textrm{MeV}$
\citep[e.g.,][]{foucart:15}.  We can
then estimate the neutrino processing timescale as
\begin{equation}
\tau_\nu(r) \approx 67.8 \, \textrm{ms} \,
\left(\frac{r}{250 \, \textrm{km}} \right)^{2}
L_{\nu_e,53}^{-1} T_{\nu_e,5}^{-1} \, ,
\end{equation}
where $r$ is the radius of the fluid element, $L_{\nu_e,53}$ is the electron
neutrino luminosity in units of $10^{53}\,\textrm{erg s}^{-1}$, and
$T_{\nu_e,5}$ is the electron neutrino spectral temperature in units of 5~MeV.
Electron antineutrinos are unlikely to contribute significantly to the neutrino
interaction timescale.  This is because in the low entropy outflows of \BHNS{}
mergers almost all protons are locked in heavy nuclei and thus have very low
neutrino capture cross-sections.

The change in $Y_e$ due to neutrino interactions can be estimated by assuming
that the tidal ejecta has a constant velocity $v$, the neutrino luminosity is
constant, electron and positron capture are unimportant, protons are locked into
heavy nuclei, and there is a finite time after merger at which neutrinos start
being emitted from the disk. With these assumptions, the evolution of $Y_e$ as
a function of radius is given by
\begin{equation}
\frac{d Y_e }{dr} = \frac{\theta(r-vt_{\nu,\text{on}})}
{v \tau_\nu(r) Y_{e,\text{eq}}} \left( Y_{e,\text{eq}} - Y_e \right),
\end{equation}
where $Y_{e,\text{eq}} = \langle Z \rangle_{\textrm{nuclei}}/\langle A
\rangle_{\textrm{nuclei}}$, and
$t_{\nu,\text{on}}$ is the time after merger at which the neutrino
luminosities reach their saturation value.

Assuming a constant average proton and neutron numbers of the heavy nuclei,
this can easily be integrated to large radius to find the final electron
fraction
\begin{align}
Y_{e,f} &\approx Y_{e,\text{eq}}
\left[1 - \exp \left(-\frac{r_0 }
{v \tau_\nu(r_0) Y_{e,\text{eq}}}\right) \right]
\nonumber\\
&\phantom{\approx}{}+ Y_{e,i} \exp \left(-\frac{r_0}
{v \tau_\nu(r_0) Y_{e,\text{eq}}}\right),
\end{align}
where $r_0 = t_{\nu,\text{on}} v $. Using the outflow velocity and neutrino
luminosities calculated in the M12-7-S9 model of \cite{foucart:14a} ($v \approx
0.25\,c$, $L_{\nu_e} \approx 10^{53} \, \textrm{erg s}^{-1}$, and
$t_{\nu,\text{on}} \approx 3 \, \textrm{ms}$) we find that the post neutrino
interaction electron fraction is $Y_{e,f} \approx 0.07$ if the $Y_{e,\text{eq}}$
is close to a half. Given that the r-process is robustly produced below $Y_e
\approx 0.25$, this suggests that neutrino interactions are much less likely to
play a significant role in determining the composition of the ejecta in \BHNS{}
mergers relative to binary \NS{} mergers, although this estimate is sensitive to
$t_{\nu,\text{on}}$ and the velocity of the outflow.

To make this more concrete, we run nucleosynthesis calculations for the M12-7-S9
model including neutrino interactions induced by a constant neutrino luminosity,
modeled as described above.  Similar results are found for the other two models
discussed in \cref{sec:binaries}.  In \cref{fig:weak_rates}, the weak
interaction rates and the electron fraction are shown for a single particle.
Because our Lagrangian trajectories start at 3 ms after the merger, the initial
density in the ejected material is below about $10^{10}\,\text{g cm}^{-3}$ and
lepton captures are dominated by neutrino captures for neutrino luminosities
above about $10^{52} \, \textrm{erg s}^{-1}$. The neutrino interaction rates
fall off as a power law in time, since this particular particle is moving away
from the merger site at constant velocity in a nearly radial direction. Other
particles can deviate from power law behavior at early times, but not strongly.
As was expected from our estimates above, the neutrino interaction timescale is
long compared to the outflow timescale and very little evolution of the electron
fraction occurs during the first 10 ms. The evolution of $Y_e$ after about 20 ms
is driven by beta-decays occurring during the r-process.

To look at the effect of weak interactions globally, the distribution of $Y_e$ in
the material ejected in model M12-7-S9 is shown in 
\cref{fig:ye_distribution} for a range of assumed neutrino luminosities. 
The GRHD simulations described in \cref{sec:binaries} include electron and
positron captures, but do not include neutrino captures. The SPH simulations
which follow the long term evolution of the ejecta include no weak interactions.
Therefore, we include weak interactions in our post-processing nucleosynthesis
calculations to assess their impact on $Y_e$.  
As we expect, the ejected material is very
neutron-rich, but becomes slightly less neutron rich with increasing electron
neutrino luminosity. The distribution of the electron fraction in the whole \NS{} is
also shown to emphasize that the ejecta in the absence of neutrinos has a
significantly lower $Y_e$ than the average $Y_e$ of a cold $1.2 \, M_\odot$ \NS{}
calculated using the LS220 \EOS{}. The beta-equilibrium value of $Y_e$ increases
with density, so that the outer layers of the NS---which comprise most of the
ejecta---have a lower electron fraction. The average electron fraction in the
ejecta is 0.053, 0.053, 0.054, 0.062, and 0.127, for neutrino luminosities of
$\{0, 0.2, 1, 5, 25\} \times 10^{52}\text{ erg s}^{-1}$.

\subsection{Nucleosynthesis and Neutrino Induced Production of the First
r-Process Peak}
\label{sec:first_peak}

\begin{figure}
\includegraphics[width=\linewidth]{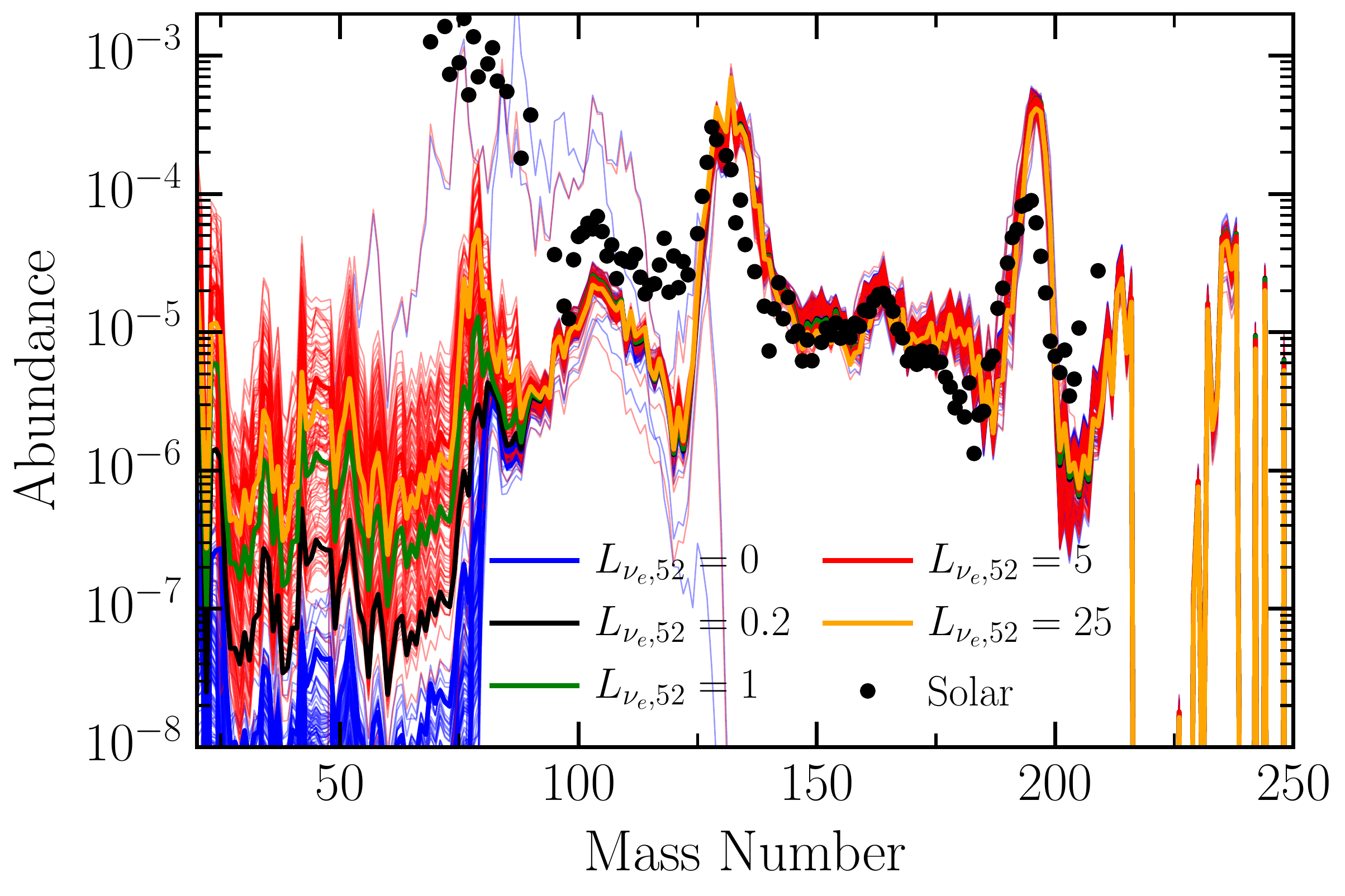}
\caption[]{Comparison of the integrated nuclear abundances in model M12-7-S9
assuming different fixed neutrino irradiation from the nascent accretion disk.
We also include the classical scaled solar abundance r-process distribution from
\cite{arlandini:99} for comparison.  \update{The abundances for a selection of
single particles from the $L_{\nu_e}=0$ and $L_{\nu_e,52} = 5$ runs are also shown as
light lines.} For all runs, we assume $L_{\bar \nu_e} = 1.5 L_{\nu_e}$.}
\label{fig:luminosity_abundances}
\end{figure}

We now consider the detailed nucleosynthesis in the ejecta of model M12-7-S9,
both with and without neutrinos. We focus on the effect neutrinos can have on
the isotopic abundances of the ejecta.
In \cref{fig:luminosity_abundances}, the integrated nucleosynthesis from
model M12-7-S9 is shown.  Since the neutrino emission from the accretion
torus formed after the \BHNS{} merger is uncertain, we calculate the final
nucleosynthetic yields of M12-7-S9 \update{assuming the range of electron neutrino luminosities
listed above.}  In all cases, the
electron antineutrino luminosity is fixed at $1.5 L_{\nu_e}$ to very
approximately account for re-leptonization of the neutrino emitting disk
\citep{foucart:15b}.  Because of the $\alpha$-effect, the results are
insensitive to the chosen electron antineutrino luminosity.  The electron
neutrino and antineutrino average energies are fixed at $12 \, \textrm{MeV}$
and $15 \, \textrm{MeV}$, respectively.  The results for the other two simulated
binary systems are similar and they are discussed briefly below. \update{We also
find that in the case of zero neutrino luminosity, the nucleosynthesis results
are not significantly altered when we use parameterized density histories with a
fixed dynamical timescale for all particles. This is not surprising, given the
low electron fraction encountered in the ejecta, which gives rise to robust
r-process nucleosynthesis across a wide range of dynamical timescales. Nevertheless,
the position history of the particles provided by our SPH code is necessary
for estimating the amount of neutrino irradiation the particles undergo.} 

In general, we confirm previous work that has shown \BHNS{} mergers dynamically
eject a large amount of r-process rich material \citep[e.g.][]{roberts:11,
korobkin:12,just:15}. Both the second and third r-process peaks are robustly produced,
independent of the neutrino luminosity. Given the low electron fractions found
in the ejecta at the start of neutron capture, robust production of the
r-process is not surprising \citep{lippuner:15}. In all of the models, reactive
flow proceeds past the third peak before neutron exhaustion occurs in the vast
majority of the simulated fluid elements and they undergo fission cycling. We
find that fission cycles occur in the ejecta and the number of cycles is weakly
dependent on the neutrino luminosity (for the luminosities considered here).
Therefore, the abundance pattern above mass number $\sim 90$ is likely to be
robust to variations in  the total neutrino luminosity and the properties of the
merging system.  In all models, the third r-process peak is over produced
relative to the second peak.  This is discussed further in 
\cref{sec:GCE}.

We find that the abundance of the first r-process peak at mass number 78 depends
on the neutrino luminosity, in contrast to the second and third peaks which are
independent of the neutrino luminosity.  Nonetheless, in all cases it is
under-produced relative to the solar abundance when normalizing to the second
and third peaks.  This first peak production is driven by low mass r-process
seed production after material falls out of \NSE{}. This material is composed of
heavy nuclei and free neutrons when strong equilibrium ceases to hold. Since the
material is still relatively close to the accretion torus a few milliseconds
after it is ejected, a significant number of electron neutrinos can be captured
by the free neutrons.  The produced protons then rapidly capture neutrons and
form deuterium, which can then capture another deuteron to form an alpha
particle. These alpha particles can then undergo a neutron-catalyzed
triple-alpha reaction, similar to what occurs in neutron-rich neutrino driven
winds \citep{delano:71, hoffman:97}, to produce low mass seed nuclei for the
r-process \citep{meyer:98}.
This non-equilibrium neutrino induced seed production creates a distinct set of
seed nuclei that can undergo neutron capture, since the seeds produced by the
\NSE{} distribution tend to be between mass 78 and 100. A large number of the
low mass seeds do not get processed past the $N = 50$, $Z = 28$ point in the
r-process path before neutron exhaustion occurs because of the long beta-decay
half lives in that region of the chart of the nuclides. Therefore, these
neutrino produced seed nuclei are responsible for producing the first peak
r-process nucleosynthesis seen in our simulations. This effect of neutrino
irradiation of the outflow is distinct from the one discussed by
\cite{wanajo:14} and \cite{goriely:15}, where the neutrino luminosities are high
enough to push the electron fraction over $\sim\!0.25$ and stop production of
the second and third peaks.

\begin{figure}
\includegraphics[width=\linewidth]{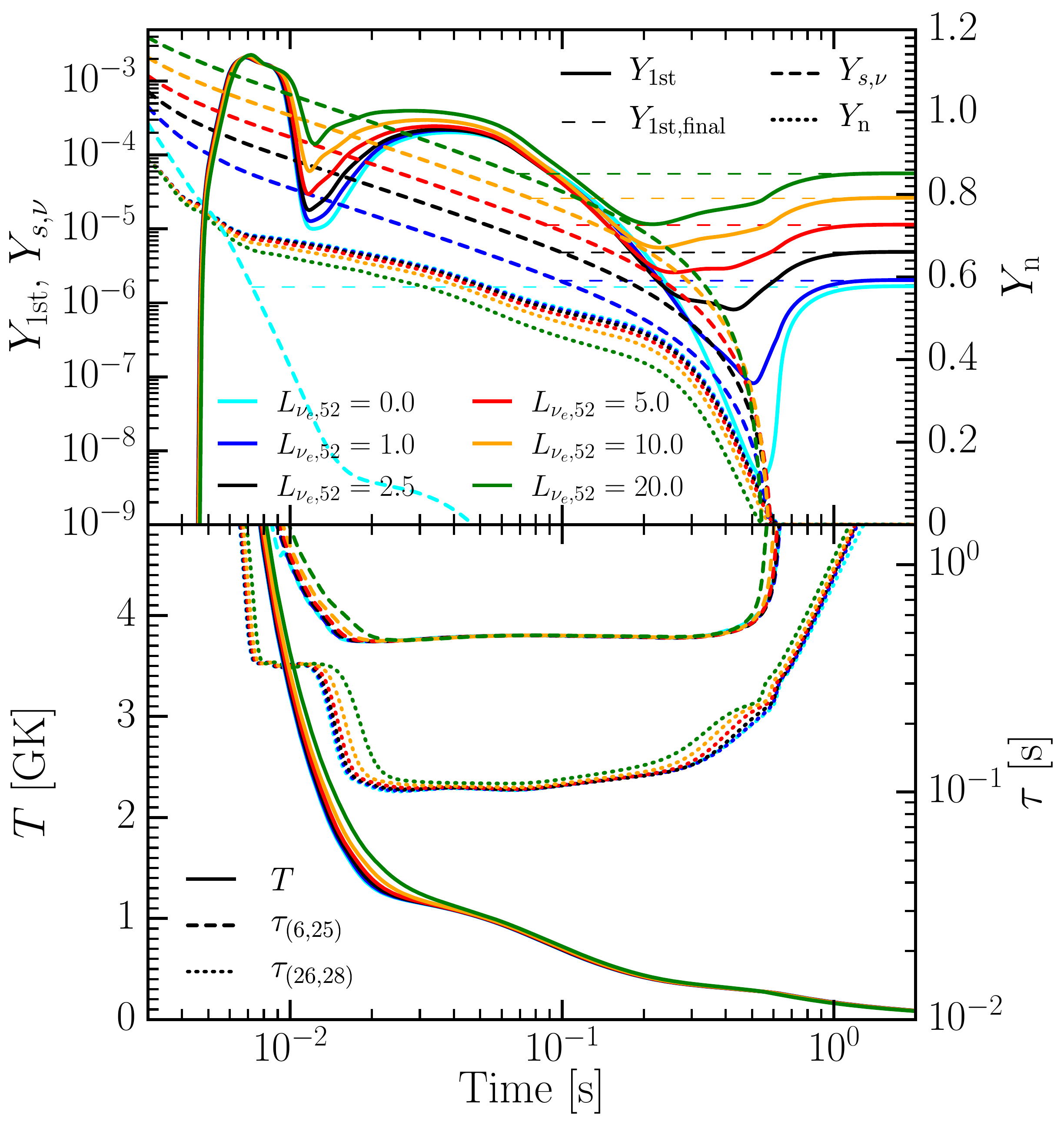}
\caption[]{Illustration of how the first r-process peak is produced by electron
neutrino captures on neutrons for a single \SPH{} particle. This \SPH{}
particle had initial $Y_e = 0.11$, initial entropy \smash{$s = 9.7\ k_B\
\text{baryon}^{-1}$}, and an asymptotic
velocity $v/c = 0.5$. {\it Top panel:} The solid lines show the abundance of
material in the first r-process peak, $Y_\text{1st}$, as a function of time
(i.e. material with
$72 \leq A \leq 79$), the dashed lines show the integrated number of protons produced
by weak interactions after time $t$ divided by six, \smash{$Y_{s,\nu} =
\int_t^{\infty} dt Y_{\textrm{n}}/6 (\lambda_{\nu_e} + \lambda_{e^+})$}, and the
dotted lines show the neutron abundance $Y_n$. $Y_{s,\nu}$ gives the number of
low mass seed nuclei produced by
neutrino interactions. The neutrino seed nuclei produced at early times are
burned past the first r-process peak, but the seed nuclei produced after the
time when $Y_{s,\nu} = Y_\text{1st,final}$ do not get burned passed the first
peak before neutrons are exhausted, and so they will end up in the first peak.
{\it Bottom panel:}
The solid lines show the temperature of the particle as a function of time, the
dashed lines show the timescale to process material to the first peak,
\smash{$\tau_{(6,25)}$}, and the dotted lines show the destruction timescale of
the first peak, \smash{$\tau_{(26,28)}$}, which are defined in the text. In this
particle, there is no significant variation with neutrino luminosity of the
temperature or the r-process path. Therefore, the two timescales do not change
with the amount of neutrino irradiation.} \label{fig:first_peak_production}
\end{figure}

\begin{figure}
\includegraphics[width=\linewidth]{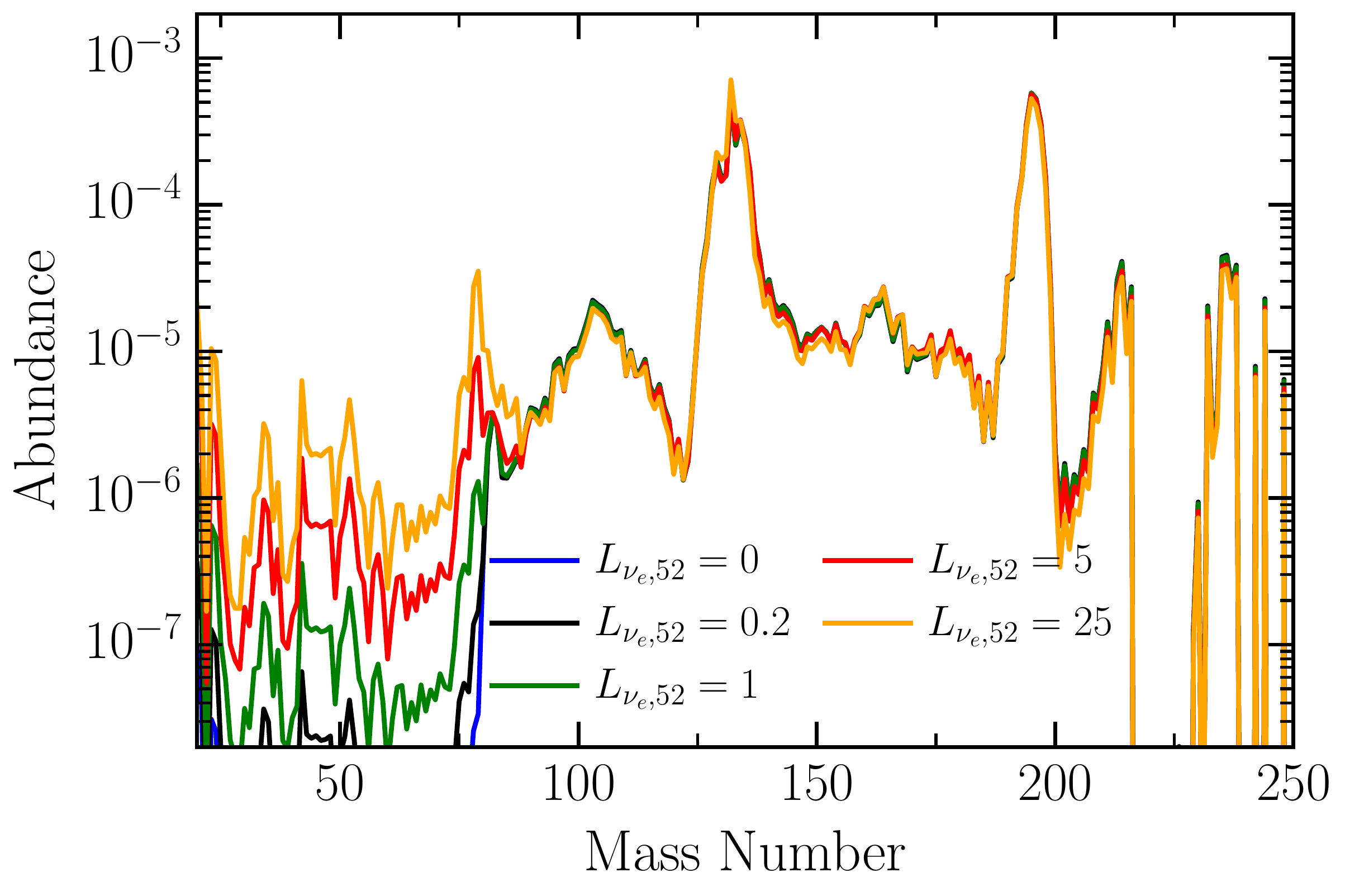}
\caption[]{Final abundances as a function of neutrino luminosity for the single Lagrangian 
particle shown in \cref{fig:first_peak_production}}
\label{fig:particle_abundances}.
\end{figure}

\begin{figure}
\includegraphics[width=\linewidth]{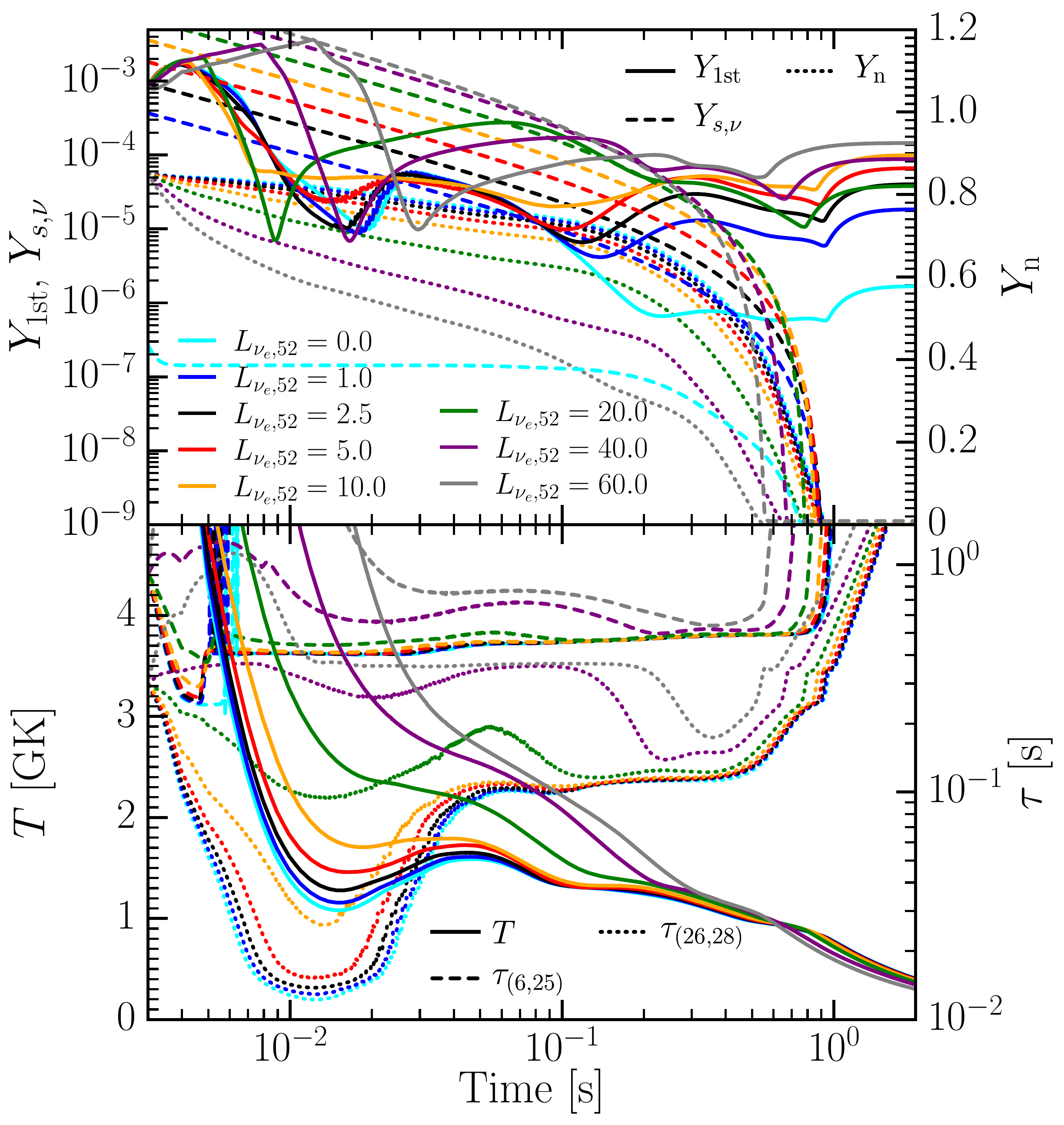}
\caption[]{The same as \cref{fig:first_peak_production}, except for
a different thermodynamic trajectory. This \SPH{} particle had initial $Y_e =
0.05$, initial entropy \smash{$s = 4.33\ k_B\ \text{baryon}^{-1}$}, and an
asymptotic velocity $v/c = 0.29$. Because of the lower velocity, lower initial
entropy, and lower $Y_e$ present in this particle relative to the particle
shown in \cref{fig:first_peak_production}, neutrino interactions significantly
alter the thermodynamic state of the material and \smash{$\tau_{(6,25)}$}. This
causes the first peak abundance to vary non-monotonically with the neutrino
luminosity.}
\label{fig:first_peak_production2}
\end{figure}

\begin{figure}
\includegraphics[width=\linewidth]{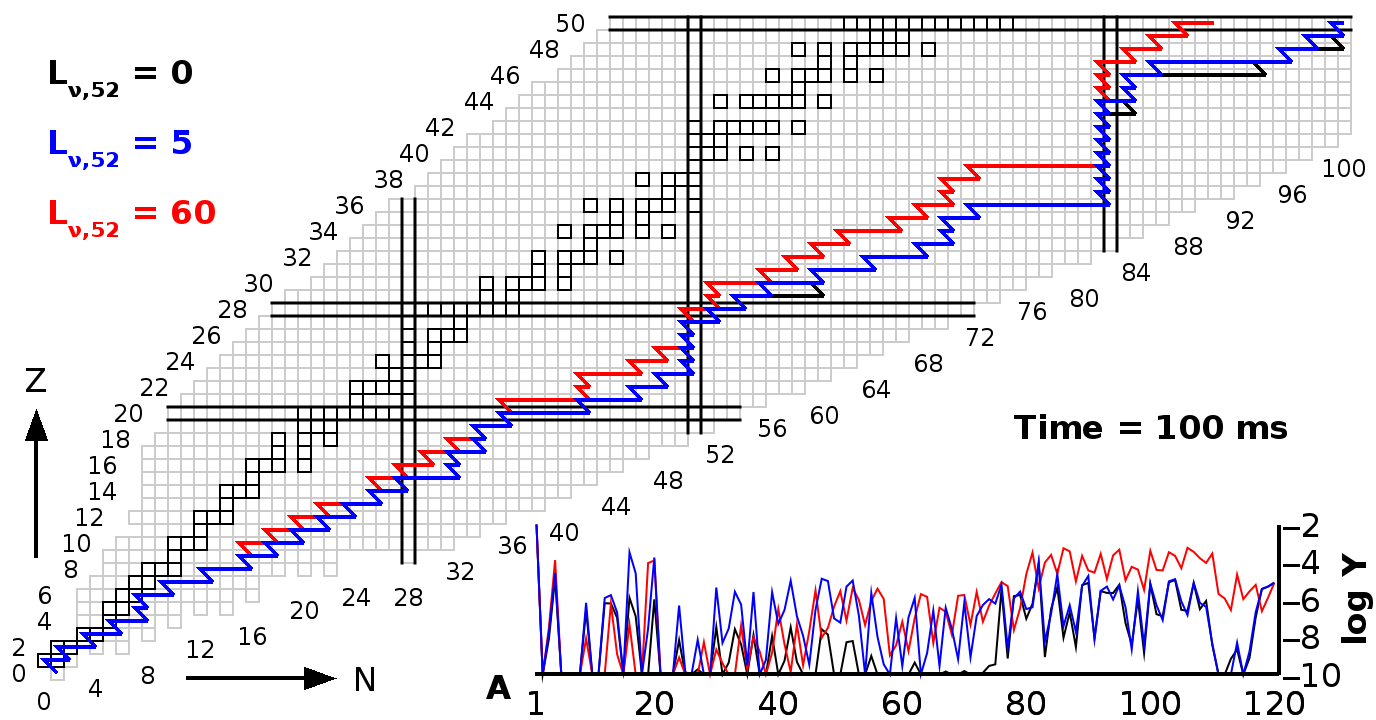}
\caption[]{The r-process path for the SPH particle shown in
\cref{fig:first_peak_production2} for different neutrino luminosities at 100 ms
into the calculation.  The inset shows the mass summed abundances at the same
time.  Notice how the path differs for different neutrino luminosities.}
\label{fig:path}
\end{figure}

\subsection{Details of the First Peak Production Mechanism}
We now consider the details of the process by which abundances in the first peak
are indirectly produced by electron neutrino captures by neutrons.
The total number fraction of heavy nuclei produced by neutrino induced seed production
can be estimated by using the results from \sref{sec:ye} as follows. Low mass
seed production proceeds via the neutron catalyzed triple alpha process, so it
takes six protons to make a seed nucleus. The rate of proton production is just
\smash{$\dot Y_e$}, so the total number of low mass seed nuclei produced by
neutrino interactions is
\begin{align}
Y_{\textrm{s},\nu} \approx& \frac{Y_{e,f} - Y_{e,i}}{6} \nonumber \\
=& \frac{Y_{e,\text{eq}} - Y_{e,i}}{6}
\left[1 - \exp \left(-\frac{r_0}
{v \tau_\nu(r_0) Y_{e,\text{eq}}}\right) \right].
\end{align}
This estimate implies that around $2\times10^{-3}$ seed nuclei per baryon are
produced by this process, assuming the neutrino luminosity is \smash{$10^{53} \,
\textrm{erg s}^{-1}$}. This number is in good agreement with the values
extracted from our nucleosynthesis calculations. Comparing this to the total
final abundance of the first peak for a single ejecta particle shown in
\cref{fig:first_peak_production}, it is clear that only about 10\% of this
material gets trapped in the first peak.

Seed nuclei indirectly produced by neutrinos are not processed past the $N = 50$
closed shell rapidly. If such rapid processing were the case, the final amount
of mass in the first peak would be set by the number of seed nuclei produced
after a time just before neutron exhaustion. To illustrate when the nuclei
trapped in the first peak are produced, we show the total number of seed nuclei
produced by neutrino interactions after time $t$
\begin{equation}
Y_{s,\nu}(t) = \frac{1}{6}\int_t^\infty Y_n (\lambda_{\nu_e} + \lambda_{e^+}) dt
\end{equation}
in \cref{fig:first_peak_production}, along with the time dependence of the first
peak abundance, $Y_\text{1st}$,  and the neutron abundance $Y_\text{n}$.
$Y_{s,\nu}$ is just the number fraction of protons produced by weak interactions
after time $t$ divided by six, since it requires six protons to produce a seed
nucleus that can capture neutrons.  Material will be processed through the first
peak on some timescale \smash{$\tau_\text{1st}$}.  Let $t_\text{ex}$ be the time
at which neutrons are exhausted and $t_\text{prod}$ be the time after which
neutrino produced seed nuclei get trapped in the first peak. Seed nuclei
produced at times earlier than $t_\text{prod} = t_\text{ex} - \tau_\text{1st}$
will be burned past the first peak, while seed nuclei produced within a time
$\tau_\text{1st}$ of neutron exhaustion will end up in the first peak. We can
estimate $\tau_\text{1st}$ by looking for solutions of $Y_{s,\nu}(t_\text{prod})
= Y_\text{1st,final}$. Inspecting \cref{fig:first_peak_production}, we find
$t_\text{prod}$ is 70 to 100~ms and $t_\text{ex}$ is 520 to 600~ms for
\smash{$L_{\nu_e,52}$} ranging from 20 to 1. Thus we estimate that
$\tau_\text{1st}$, the time it takes for seed nuclei to be processed to the $N =
50$ closed shell of the first peak, is between 450 and 500~ms for this
particular fluid element. \update{For reference, the final abundances of this particle 
are shown in \cref{fig:particle_abundances}}.

We now attempt to explain what sets this timescale.  Assuming \update{the
waiting point approximation \citep[c.f.][]{kratz:93}},
the timescale to go from charge $Z_1$ to charge $Z_2$ is given by
\begin{equation}
\tau_{(Z_1,Z_2)}(t) = \sum_{Z=Z_1}^{Z_2}
\frac{\sum_N Y_{(Z,N)}} {\sum_N Y_{(Z,N)} \tau^{-1}_{\beta^{-},(Z,N)}}.
\end{equation}
Here, \smash{$\tau^{-1}_{\beta^{-}, (Z,N)}$} is the beta-decay timescale of a
nucleus with $N$ neutrons and $Z$ protons. When \smash{$(n,\gamma)$} reactions
are in equilibrium with \smash{$(\gamma, n)$} reactions---such that
\smash{$\mu_n + \mu_{(Z,N)} = \mu_{(Z,N+1)}$}---these timescales are only
functions of the density, temperature, and neutron abundance, i.e.\
\smash{$\tau_{(Z_1,Z_2)} = \tau_{(Z_1,Z_2)}(\rho, T, Y_n)$}.  This is often the
case at the high temperatures encountered during r-process nucleosynthesis in
these outflows, but the equilibrium can start to break down at lower
temperatures \citep{kratz:93}.  This makes it clear that changing the temperature and electron
fraction of a particular fluid element can change the path of the r-process and
alter the time it takes material to be processed from one charge number to
another. The quantities \smash{$\tau_{(6,25)}$} and \smash{$\tau_{(26,28)}$} are
shown in the bottom panel of \cref{fig:first_peak_production}. Note that
\smash{$\tau_{(6,25)} + \tau_{(26,28)}$} is approximately the time it takes a
seed nucleus to get to the start of the first peak and then get processed
through the first peak, which we called
$\tau_\text{1st}$ above. We see from the bottom panel of
\cref{fig:first_peak_production} that \smash{$\tau_{(6,25)}$} is constant
throughout the period during which the r-process is occurring and its value is
in good agreement with our 450 to 500~ms estimate for \smash{$\tau_\text{1st}$}.
Because this timescale is determined by beta decay, the final first peak
abundance goes linearly with the neutrino luminosity.  The lifetimes of isotopes
along the $N=50$ closed shell are 40~ms, 110~ms, and 110~ms, for the reactions
\smash{$^{76}$Fe($\beta^-,n)^{75}$Co}, \smash{$^{77}$Co($\beta^-,n)^{76}$Ni},
and \smash{$^{78}$Ni($\beta^-)^{78}$Cu}.  These are consistent with the
\smash{$\tau_{(26,28)} \approx 100 \ \textrm{ms}$} we find in
\cref{fig:first_peak_production}. We also note that at around 600~ms into the
calculation---which is after neutron exhaustion---there is a further increase in
the first peak abundance.  This is driven by the reaction
\smash{$^{80}$Ni($\beta^-,n)^{79}$Cu}. Significant production of
\smash{$^{80}$Ni} occurs just before neutron freeze-out and it has a half-life
of 175~ms.  

This suggests that the neutrino flux between times $t_1 = t_\textrm{ex} -
\tau_{(6,25)} - \tau_{(26,28)}$ and  $t_2 = t_\textrm{ex} - \tau_{(6,25)}$ will
determine the amount of neutrino induced first peak production that occurs.
Seed nuclei produced before time $t_1$ will get beyond the first peak before
neutrons are exhausted, while seed nuclei produced after time $t_2$ will not
reach the first peak before neutron exhaustion occurs.  Therefore, the important
quantity for understanding neutrino induced production of the first peak will be
the neutrino luminosity centered at a time around 70~ms after merger, within a
window of around 100~ms.  

We have shown that the production of first peak nuclei goes linearly with the
electron neutrino luminosity for the specific Lagrangian particle shown in
\cref{fig:first_peak_production}, but \cref{fig:luminosity_abundances} shows
that production of the first peak appears to saturate at luminosities above
$\sim 5\times 10^{52} \, \textrm{erg s}^{-1}$.  Below this luminosity, the
dependence of first peak production on luminosity is approximately linear as
expected.  In \cref{fig:first_peak_production2}, we show a different Lagrangian
particle that exhibits non-monotonic behavior of the first peak abundance with
the neutrino luminosity.  The first peak abundance increases at low luminosity,
decreases with luminosity around $L_{\nu_e} = 10^{53} \, \textrm{erg s}^{-1}$, and
then increases with luminosity again.  This particle has a lower asymptotic
velocity than the particle shown in \cref{fig:first_peak_production} and
therefore experiences more neutrino irradiation. Additionally, it has lower
initial entropy and $Y_e$, which means neutrino interactions can have a larger
effect on its thermodynamic state.

The lower panel of \cref{fig:first_peak_production2} clearly shows that
neutrinos significantly alter the thermodynamic state of the considered particle
and that the low mass r-process path is shifted by the inclusion of neutrino
interactions.  \update{In particular, \smash{$\tau_{(26,28)}$} and
\smash{$\tau_{(6,25)}$} increase with the neutrino luminosity. 
\cref{fig:path} shows how the r-process path varies with the neutrino luminosity
100 ms into the calculation, giving rise to the processsing timescales
dependence on neutrino luminosity.}  The total number of seeds increases with
initial $Y_e$ and temperature, corresponding to larger neutrino luminosities in
this fluid element. Additionally, increasing the entropy of the outflow reduces
the rate at which material can bypass the $A=8$ stability gap.  The large
difference in the first peak processing timescale, \smash{$\tau_{(26,28)}$},
seen in \cref{fig:first_peak_production2} is due to the r-process path shifting
from being far beyond the $N=50$, $Z=28$ closed shells at lower temperatures
(and lower neutrino luminosities) to proceeding through closed shells at higher
temperatures (and higher neutrino luminosities).  This significantly alters how
first peak nuclei are produced throughout the calculation and breaks the linear
dependence on the neutrino luminosity.

Even in the absence of neutrinos, there is some production of first peak nuclei.
As we have mentioned, this material is produced by fission of heavy nuclei.
Since we are employing symmetric fission fragment distributions, it is likely
that more realistic fission fragment distributions will result in a broader
distribution of fission daughters and more material being left behind in this
region. Nonetheless, it seems likely that there will be at least some production
of the first peak even in the very neutron-rich outflows of \BHNS{} mergers, as
long as neutrino luminosities from the post merger remnant are above about
$10^{52} \, \textrm{erg s}^{-1}$ within a hundred milliseconds of the merger. We
also emphasize that neutrino induced production of the first peak does not
produce enough material in our models to agree with the solar r-process
abundances when they are normalized to the second peak.  Instead, the abundance
is around an order of magnitude too low.

\subsection{Isotopic and Elemental Abundances, Galactic Chemical Evolution, and
Low Metallicity Halo Stars}
\label{sec:GCE}

\begin{figure}
\includegraphics[width=\linewidth]{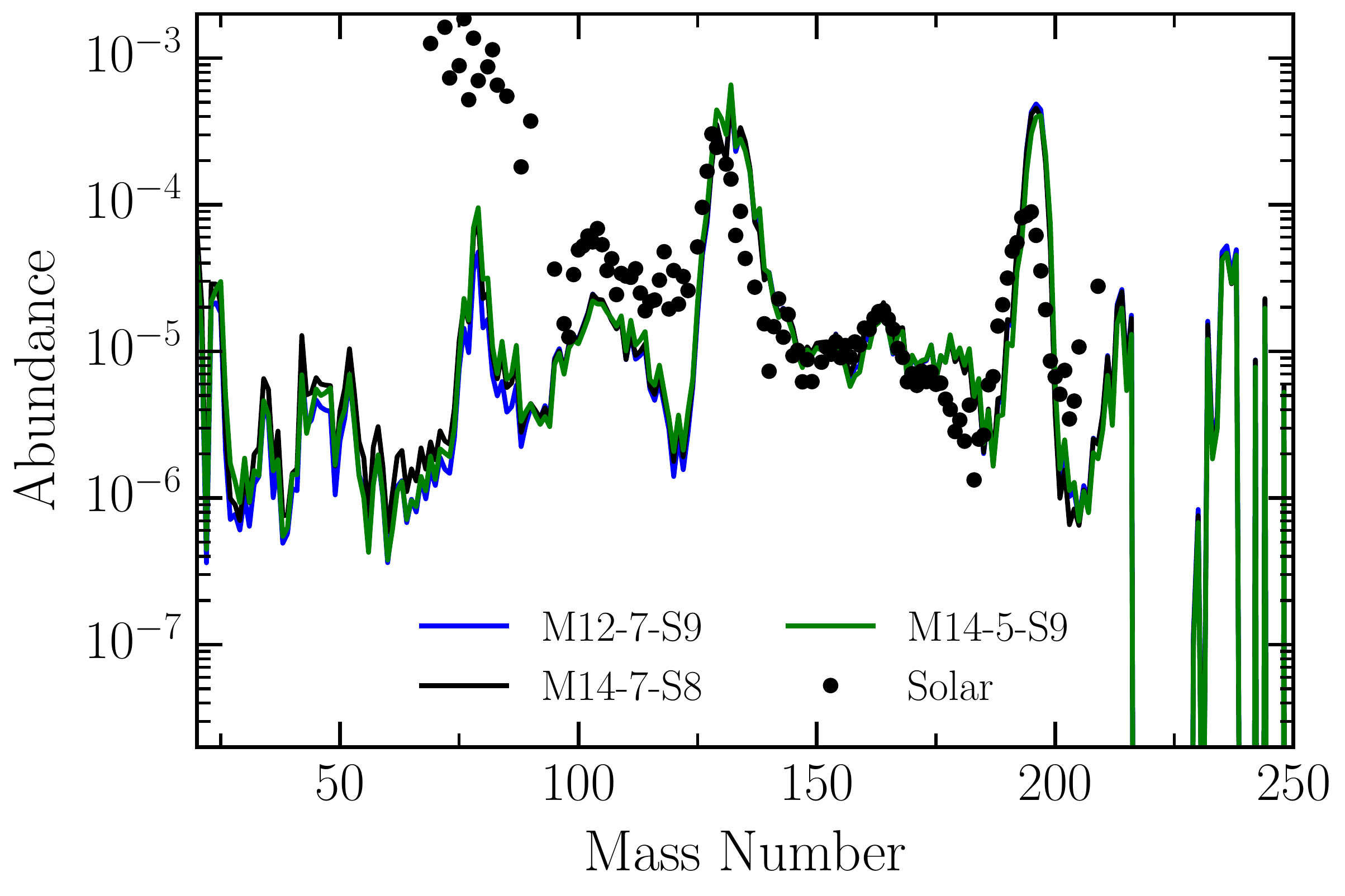}
\caption[]{The integrated nuclear abundances of the dynamical ejecta in the
models M12-7-S9, M14-7-S8, and M14-5-S9 assuming an electron neutrino luminosity
of $10^{53} \, \textrm{erg s}^{-1}$. The low electron fraction of all the ejecta
results in production of a robust r-process, independent of the dynamics of the
merger and parameters of the binary system.}
\label{fig:integrated_abundances}
\end{figure}

In \cref{fig:integrated_abundances}, the integrated abundances from the models
M12-7-S9, M14-7-S8, and M14-5-S9 are shown for a fixed neutrino luminosity of
$10^{53} \, \textrm{erg s}^{-1}$ (and $L_{\bar \nu_e} = 1.5 L_{\nu_e}$). Clearly,
there is little discernible difference between the predicted nucleosynthesis
from these models. The electron fraction in almost all of the ejecta in all
three models is below the threshold for fission cycling to occur
\citep{lippuner:15} and the entropy of the ejecta is quite low. These are
conditions that result in a second and third peak nuclear abundance pattern that
is quite insensitive to the detailed properties of the ejecta
\citep[e.g.]{lippuner:15}. Interestingly, the neutrino produced first peak is
also very insensitive to the binary parameters if the same neutrino luminosities
are assumed. A priori, it would seem that different dynamics during the merger
could give rise to different dynamics of the ejecta and alter the radius at
which neutron exhaustion occurs.  Of course, the important parameter will be the
velocity of the ejecta. Larger velocities will result in lower local neutrino
densities around the time that neutrons are exhausted in the ejected material.

Although there is reasonable qualitative agreement with the solar r-process
abundance pattern above $A \sim 100$ in all of our models, there are significant
quantitative differences. Most clearly, the third peak is significantly over
produced and has an offset with respect to the observed solar pattern.
Uncertainties in the ejecta abundance pattern can come from two sources,
uncertainties in the properties of the ejecta and uncertainties in the nuclear
data that serves as input for our nucleosynthesis calculations. Given how
robust the r-process pattern in our models is above $A\sim100$ to variations in
the binary parameters and neutrino irradiation, it seems unlikely---{\it for
our chosen nuclear data}---that \BHNS{} mergers can make a pattern that agrees
exactly with the r-process pattern found in the Sun and low metallicity halo
stars. Strictly following this argument to its conclusion, \BHNS{} mergers
would be ruled out as the dominant contributor to the galactic chemical
evolution of r-process elements. This would put a significant constraint on the
combined merger rate, the \BH{} spin distribution, and the \BH{} mass
distribution in these binaries \citep{bauswein:14}. Of course, it is easy to
imagine scenarios where the galactic r-process nucleosynthesis is produced by
multiple types of events, so strongly ruling out a single r-process production
channel on the basis of inexact agreement with the solar pattern seems premature
at best.

The second possible source of uncertainty in our results is the input nuclear
physics data. The nuclear masses, beta-decay rates, neutron capture rates,
\update{fission barrier heights,} and
fission fragment distributions in the r-process path, which lies far from
nuclear stability, have, on the whole, not been experimentally determined but
are instead determined from models \update{that are in part constrained by data
from nuclei closer to stability
\citep[e.g.][]{moeller:97, goriely:09}}. For instance, different nuclear mass models can give
significantly different abundance patterns for the same thermodynamic histories
\citep[e.g.][]{arcones:11, martin:15, mendoza-temis:15}. By varying masses within a particular mass model
within the expected uncertainty, \cite{mumpower:15} have shown that the
uncertainties in the final r-process abundance pattern solely due to nuclear
physics uncertainties can be as a large as a factor of ten. \update{We have also not
included neutrino-induced fission in our nuclear network \citep{qian:02, kolbe:04}, which could change
the nuclei that undergo fission and potentially alter the low mass r-process
pattern. Nevertheless, the neutrino irradiation in these outflows is rather weak
so it would be suprising if neutrino induced fission drastically changed our
results.} Therefore, given the
level of agreement we find with the solar r-process isotopic abundance pattern,
our results seem wholly consistent with \BHNS{} mergers contributing to the
galactic budget of heavy r-process nuclei.

\begin{figure}
\includegraphics[width=\linewidth]{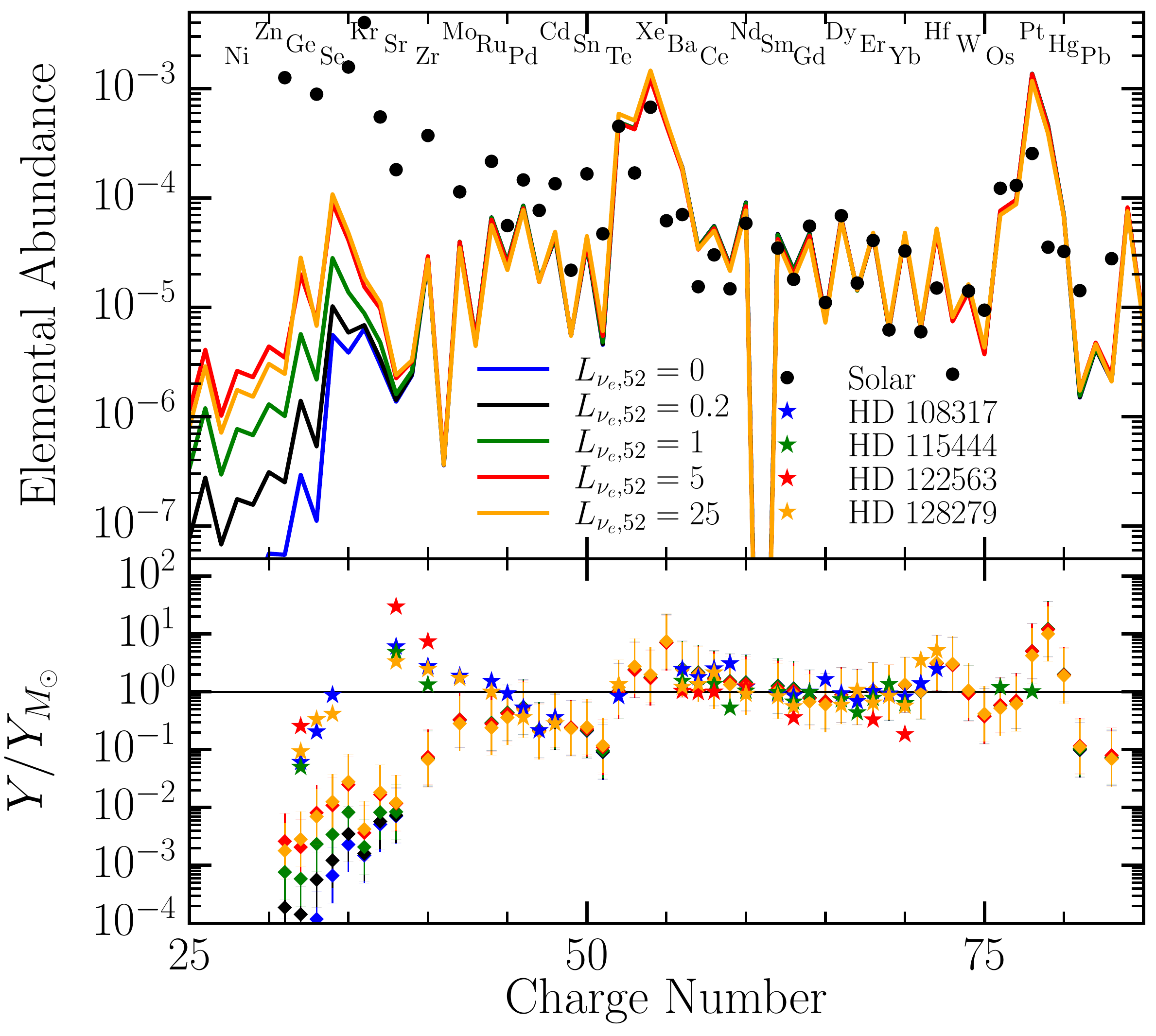}
\caption[]{{\it Top Panel:} The integrated elemental abundances in the ejecta of
model M12-7-S9 for a range of imposed neutrino luminosities compared to
arbitrarily scaled solar r-process abundances. We include the name of every
other element to guide the eye. {\it Bottom Panel:} Ratio of the calculated
abundances to the solar abundances \citep[from][]{arlandini:99}. In the lower
panel, factor of three error bars are included on our calculated abundance
patterns to approximately account for errors in the input nuclear physics. The
solar r-process abundances have been transformed by an overall scaling factor to
minimize the residuals in log space in the region $40<Z<81$. We also include
abundances from the low metallicity halo stars HD 115444 and 122563
\citep{westin:00} and HD 108317 and 128279 \citep{roederer:12, roederer:14},
with [Fe/H] of $-3.0$, $-2.7$, $-2.5$, and $-2.5$, respectively. The observational
error bars are smaller than the plotted symbols and the abundances patterns have
been scaled by a factor to minimize the deviation from the solar r-process in
the range $40<Z<81$.}
\label{fig:integrated_Zabundances}
\end{figure}

For confrontation with observations of abundances in low metallicity halo stars,
it is more instructive to examine the {\it elemental} abundance pattern of the
ejecta (i.e. $Y_{\rm Z} = \sum_i \delta_{Z_i, {\rm Z}} Y_i$)\update{,
since only elemental abundances are easily determined in these stars}. In
\cref{fig:integrated_Zabundances}, we show the final elemental abundances in the
ejecta of model M12-7-S9 for a variety of assumed neutrino luminosities. In the
region $51<Z<81$, most of the abundances agree with the solar pattern to within
a factor of three. The notable exceptions are gold and platinum in the third
peak and cesium in the second peak, all of which are overproduced by a factor of
around ten. Although the agreement is not perfect, our patterns above Mo are
within the errors due to uncertainties in the nuclear physics input
\citep{mumpower:15}.

For elements below Mo, there are a number of primary nucleosynthesis
processes that can contribute to these abundances, even in low metallicity stars
\citep{travaglio:04, montes:07, qian:08, arcones:11b, hansen:14}. Therefore, it is hard to rule out our
models because they fail to produce certain abundances below Mo. As can be seen
in the bottom panel of \cref{fig:integrated_Zabundances}, the abundances
below Mo are significantly underproduced relative to solar in all of our
models. Nevertheless, neutrino induced r-process seed production brings these
abundances up much closer to the solar value.

Since many processes contribute to the solar abundances, it is more reasonable
to compare our abundances with those measured in the atmospheres of low
metallicity halo stars. In \cref{fig:integrated_Zabundances}, we show the
abundances of four low metallicity halo stars with sub-solar Ge abundances. The
abundance data are taken from \cite{westin:00}, \cite{roederer:12}, and
\cite{roederer:14} for the stars HD 115444, 122563, 108317, and 128279, which
have metallicities of $[\textrm{Fe/H}]=-3.0$, $-2.7$, $-2.5$, and $-2.5$, respectively.
These metallicities are low enough such that s-process contamination from low
mass stars is highly unlikely \citep[e.g.]{simmerer:04}. Nonetheless, it is possible
that the s-process in massive stars may have contributed to these abundances
\citep[e.g.,][]{woosley:02}.

\update{We can compare our predicted abundances with observed abundances of Ge, As,
and Se in low metallicity halo stars}. For these elements, neutrino induced r-process seed production of the
first peak brings the abundances of Ge, As, and Se closer to agreement with the
observed values for larger values of the electron neutrino luminosity, even
though the abundances are still small by close to a factor of ten.  Given the
significant uncertaties in the level of neutrino irradiation and the properties
of nuclei far from stability near and below the first peak, the actual amount of
first peak material created by neutrino induced seed production is highly
uncertain.  If first peak production could be increased by around an order of
magnitude, certain low metallicity halo star abundance patterns above Zn can
possibly be explained by only invoking nucleosynthesis in the dynamical ejecta
of \BHNS{} mergers, although there are clearly a number of other possible ways
to produce the first peak in the material ejected from the disk
\cite[e.g.][]{just:15}.  This contrasts with the results of \cite{wanajo:14} for
binary \NS{} mergers, where they find that the first peak nuclei are produced
with a ratio to the second and third peak that is close to the ratio observed in
the sun and therefore are over producing the first peak by about a factor of ten
relative to the low metallicity halo star abundances.
In contrast to Ge, As, and Se, all the \BHNS{} merger outflows underproduce Sr
by orders of magnitude and the final abundances of Sr and Zr are insensitive to
the level of neutrino irradiation. \remove{This discrepancy could be due in part to our
use of symmetric fission fragment distributions, another process providing these
elements (such as neutrino driven winds in core collapse supernovae;
), or it could signal problems for a \BHNS{} merger production
scenario for the r-process.}

\section{Conclusions}

In this work, we have considered nucleosynthesis in the ejecta of \BHNS{}
mergers including the effect of neutrino interactions. Starting from general
relativistic hydrodynamic simulations of \BHNS{} mergers with reasonably high
\BH{} spins, we have extracted the unbound material and mapped it into a
Newtonian \SPH{} code, \StarSmasher.  We then evolved the ejecta over a long
enough time to reach homologous expansion. Using the Lagrangian histories of
the \SPH{} particles, we then performed post-processing calculations of
nucleosynthesis in the neutron-rich ejecta with \SkyNet. In particular, we
focused on the influence of neutrinos on the nucleosynthesis in these outflows
by parameterizing the neutrino luminosity coming from the disk.

As is expected from previous work \citep[e.g.][]{lattimer:77, freiburghaus:99, roberts:11,
korobkin:12, foucart:14a, bauswein:14}, we find that the second and third r-process peaks are
robustly produced in the outflow. In contrast to the case of \NSNS{}  mergers
\citep{wanajo:14, goriely:15, sekiguchi:15}, we find that---for reasonable
luminosities of $L_{\nu_e} = \{0, 0.2, 1, 5, 25\} \times 10^{52} \, \textrm{erg s}^{-1}$---neutrinos
are unable to significantly shift the distribution of the pre-r-process electron
fraction in the ejecta and almost all of the ejected mass elements produce the
full r-process. This is due to the rapid outflow timescales of the tidal ejecta
and the relatively low expected neutrino luminosities of the disk formed around
the remnant \BH{}. Additionally, we find that there are negligible differences
between the nucleosynthesis calculated for different \BHNS{} binary parameters.

Although weak interactions have no gross effect on the electron fraction
distribution, we find they do alter nucleosynthesis in more subtle ways. Once
r-process neutron captures have begun, electron neutrino captures can turn a
small fraction of the neutrons into protons \update{due to the $\alpha$-effect
\citep{fuller:95}. These protons form alpha particles
which rapidly combine to form $^{12}$C. This provides a source of low mass seed
nuclei for the r-process \citep{meyer:98}} and results in nuclear flow below mass 78 that is not
present \update{in low entropy outflows} when neutrinos are neglected. In the few hundred milliseconds before
neutron exhaustion, this flow can build up a significant amount of material in
the first r-process peak at mass 78. \remove{To the best of our knowledge, this is the
first time this process has been described.} \update{This process} is likely to operate both in the
dynamical ejecta of \BHNS{} mergers and binary \NS{} mergers, as well as in
material that is ejected from the disks left over in these events, which we did
not consider in this work.

Although this neutrino induced seed production produces the first peak, we find
that it does not produce enough material to bring the final abundance
distributions into agreement with the solar r-process distribution below mass
80. Nonetheless, this first peak production will be interesting when
comparing with the abundance patterns of low metallicity halo stars. In
particular, it can significantly increase the abundance of the elements Ge, As,
and Se \update{present in these outflows}.

Further work is required to determine how robust neutrino induced
r-process seed production is to variations in beta-decay lifetimes and neutron
capture rates around the $N=50$ closed shell. Additionally, employing more
realistic fission fragment distributions than the ones employed here may also
effect the abundances found just above the first r-process peak. Of course,
this process is sensitive to the electron neutrino flux in the outflow.
Here, we have chosen to parameterize the neutrino luminosity. Therefore, better
models of the electron neutrino irradiation of the outflow are required to
determine if neutrino induced r-process seed production is robust in
nature. The results are likely to be somewhat sensitive to the beta-decay rates
away from stability below $Z=28$, so better measurements of decays along the
r-process path could shed significant light on whether or not this process
is important in the ejecta of \BHNS{} mergers.  It will also be interesting to
investigate if this process operates in other environments with low $Y_e$
outflows, namely the ejecta of binary \NS{} mergers and material ejected from
disks formed after compact object mergers.

\section*{Acknowledgements}
LR acknowledges Yongzhong Qian and Projjwal Banerjee for useful discussions
relating to this work.

Support for this work was provided by NASA through Einstein
Postdoctoral Fellowship grants numbered PF3-140114 (LR) and PF4-150122
(FF) awarded by the Chandra X-ray Center, which is operated by the
Smithsonian Astrophysical Observatory for NASA under contract
NAS8-03060. JL and CDO are partially supported by the NSF under award
Nos.\ TCAN AST-1333520, CAREER PHY-1151197, and AST-1205732, and by
the Sherman Fairchild Foundation.  JCL is supported by NSF grant
number AST-1313091. This work also benefitted from NSF support through
award No. PHY-1430152 (JINA Center for the Evolution of the Elements).
M.D.D. acknowledges support through NSF Grant PHY-1402916.


\begin{thebibliography}{}
\expandafter\ifx\csname natexlab\endcsname\relax\def\natexlab#1{#1}\fi

\bibitem[{{Aasi} {et~al.}(2015){Aasi}, {Abbott}, {Abbott}, {Abbott},
  {Abernathy}, {Ackley}, {Adams}, {Adams}, {Addesso}, \& et~al.}]{aligo:15}
{Aasi}, J., {Abbott}, B.~P., {Abbott}, R., {et~al.} 2015, \cqg, 32, 074001,
  \href{http://doi.org/10.1088/0264-9381/32/7/074001}{doi:10.1088/0264-9381/32/7/074001}

\bibitem[{{Abadie} {et~al.}(2010){Abadie}, {Abbott}, {Abbott}, {Abernathy},
  {Accadia}, {Acernese}, {Adams}, {Adhikari}, {Ajith}, {Allen}, \&
  et~al.}]{abadie:10}
{Abadie}, J., {Abbott}, B.~P., {Abbott}, R., {et~al.} 2010, \cqg, 27, 173001,
  \href{http://arxiv.org/abs/1003.2480}{arXiv:astro-ph.HE/1003.2480}

\bibitem[{{Acernese} {et~al.}(2015){Acernese}, {Agathos}, {Agatsuma}, {Aisa},
  {Allemandou}, {Allocca}, {Amarni}, {Astone}, {Balestri}, {Ballardin}, \&
  et~al.}]{avirgo:15}
{Acernese}, F., {Agathos}, M., {Agatsuma}, K., {et~al.} 2015, \cqg, 32, 024001,
  \href{http://arxiv.org/abs/1408.3978}{arXiv:gr-qc/1408.3978}

\bibitem[{{Antoniadis} {et~al.}(2013){Antoniadis}, {Freire}, {Wex}, {Tauris},
  {Lynch}, {van Kerkwijk}, {Kramer}, {Bassa}, {Dhillon}, {Driebe}, {Hessels},
  {Kaspi}, {Kondratiev}, {Langer}, {Marsh}, {McLaughlin}, {Pennucci}, {Ransom},
  {Stairs}, {van Leeuwen}, {Verbiest}, \& {Whelan}}]{antoniadis:13}
{Antoniadis}, J., {Freire}, P.~C.~C., {Wex}, N., {et~al.} 2013, Science, 340,
  448, \href{http://arxiv.org/abs/1304.6875}{arXiv:astro-ph.HE/1304.6875}

\bibitem[{{Arcones} \& {Mart{\'{\i}}nez-Pinedo}(2011)}]{arcones:11}
{Arcones}, A., \& {Mart{\'{\i}}nez-Pinedo}, G. 2011, \prc, 83, 045809,
  \href{http://arxiv.org/abs/1008.3890}{arXiv:astro-ph.SR/1008.3890}


\bibitem[{{Arcones} \& {Montes}(2011)}]{arcones:11b}
{Arcones}, A., \& {Montes}, F. 2011, \apj, 731, 5,
  \href{http://arxiv.org/abs/1007.1275}{arXiv:astro-ph.GA/1007.1275}

\bibitem[{{Arcones} \& {Thielemann}(2013)}]{arcones:13}
{Arcones}, A., \& {Thielemann}, F.-K. 2013, Journal of Physics G Nuclear
  Physics, 40, 013201,
  \href{http://arxiv.org/abs/1207.2527}{arXiv:astro-ph.SR/1207.2527}

\bibitem[{{Argast} {et~al.}(2004){Argast}, {Samland}, {Thielemann}, \&
  {Qian}}]{argast:04}
{Argast}, D., {Samland}, M., {Thielemann}, F.-K., \& {Qian}, Y.-Z. 2004, \aap,
  416, 997,
  \href{http://arxiv.org/abs/astro-ph/0309237}{arXiv:astro-ph/0309237}

\bibitem[{{Arlandini} {et~al.}(1999){Arlandini}, {K{\"a}ppeler}, {Wisshak},
  {Gallino}, {Lugaro}, {Busso}, \& {Straniero}}]{arlandini:99}
{Arlandini}, C., {K{\"a}ppeler}, F., {Wisshak}, K., {et~al.} 1999, \apj, 525,
  886, \href{http://arxiv.org/abs/astro-ph/9906266}{arXiv:astro-ph/9906266}

\bibitem[{{Balsara}(1995)}]{Balsara1995}
{Balsara}, D.~S. 1995, Journal of Computational Physics, 121, 357,
  \href{http://doi.org/10.1016/S0021-9991(95)90221-X}{doi:10.1016/S0021-9991(95)90221-X}

\bibitem[{{Bauswein} {et~al.}(2014{\natexlab{a}}){Bauswein}, {Ardevol
  Pulpillo}, {Janka}, \& {Goriely}}]{bauswein:14b}
{Bauswein}, A., {Ardevol Pulpillo}, R., {Janka}, H.-T., \& {Goriely}, S.
  2014{\natexlab{a}}, \apjl, 795, L9,
  \href{http://arxiv.org/abs/1408.1783}{arXiv:astro-ph.SR/1408.1783}

\bibitem[{{Bauswein} {et~al.}(2014{\natexlab{b}}){Bauswein}, {Stergioulas}, \&
  {Janka}}]{bauswein:14}
{Bauswein}, A., {Stergioulas}, N., \& {Janka}, H.-T. 2014{\natexlab{b}}, \prd,
  90, 023002,
  \href{http://arxiv.org/abs/1403.5301}{arXiv:astro-ph.SR/1403.5301}

\bibitem[{{Berger} {et~al.}(2013){Berger}, {Fong}, \& {Chornock}}]{berger:13}
{Berger}, E., {Fong}, W., \& {Chornock}, R. 2013, \apjl, 774, L23,
  \href{http://arxiv.org/abs/1306.3960}{arXiv:astro-ph.HE/1306.3960}

\bibitem[{Burbidge {et~al.}(1957)Burbidge, Burbidge, Fowler, \&
  Hoyle}]{burbidge:57}
Burbidge, E.~M., Burbidge, G.~R., Fowler, W.~A., \& Hoyle, F. 1957, Rev. Mod.
  Phys., 29, 547,
  \href{http://doi.org/10.1103/RevModPhys.29.547}{doi:10.1103/RevModPhys.29.547}


\bibitem[Delano \& Cameron(1971)]{delano:71} Delano, M.~D., \&
Cameron, A.~G.~W.\ 1971, \apss, 10, 203 

\bibitem[{{Deaton} {et~al.}(2013){Deaton}, {Duez}, {Foucart}, {O'Connor},
  {Ott}, {Kidder}, {Muhlberger}, {Scheel}, \& {Szilagyi}}]{deaton:13}
{Deaton}, M.~B., {Duez}, M.~D., {Foucart}, F., {et~al.} 2013, \apj, 776, 47,
  \href{http://arxiv.org/abs/1304.3384}{arXiv:astro-ph.HE/1304.3384}

\bibitem[{{Demorest} {et~al.}(2010){Demorest}, {Pennucci}, {Ransom}, {Roberts},
  \& {Hessels}}]{demorest:10b}
{Demorest}, P.~B., {Pennucci}, T., {Ransom}, S.~M., {Roberts}, M.~S.~E., \&
  {Hessels}, J.~W.~T. 2010, \nat, 467, 1081,
  \href{http://arxiv.org/abs/1010.5788}{arXiv:astro-ph.HE/1010.5788}

\bibitem[{{Farr} {et~al.}(2011){Farr}, {Sravan}, {Cantrell}, {Kreidberg},
  {Bailyn}, {Mandel}, \& {Kalogera}}]{farr:11}
{Farr}, W.~M., {Sravan}, N., {Cantrell}, A., {et~al.} 2011, \apj, 741, 103,
  \href{http://arxiv.org/abs/1011.1459}{arXiv:astro-ph.GA/1011.1459}

\bibitem[{{Fischer} {et~al.}(2010){Fischer}, {Whitehouse}, {Mezzacappa},
  {Thielemann}, \& {Liebend{\"o}rfer}}]{fischer:10}
{Fischer}, T., {Whitehouse}, S.~C., {Mezzacappa}, A., {Thielemann}, F.-K., \&
  {Liebend{\"o}rfer}, M. 2010, \aap, 517, A80,
  \href{http://arxiv.org/abs/0908.1871}{arXiv:astro-ph.HE/0908.1871}

\bibitem[{{Foucart}(2012)}]{foucart:12dm}
{Foucart}, F. 2012, \prd, 86, 124007,
  \href{http://arxiv.org/abs/1207.6304}{arXiv:astro-ph.HE/1207.6304}

\bibitem[{{Foucart} {et~al.}(2013){Foucart}, {Deaton}, {Duez}, {Kidder},
  {MacDonald}, {Ott}, {Pfeiffer}, {Scheel}, {Szilagyi}, \&
  {Teukolsky}}]{foucart:13}
{Foucart}, F., {Deaton}, M.~B., {Duez}, M.~D., {et~al.} 2013, \prd, 87, 084006,
  \href{http://arxiv.org/abs/1212.4810}{arXiv:gr-qc/1212.4810}

\bibitem[{Foucart {et~al.}(2014)Foucart, Deaton, Duez, {O'Connor}, Ott, Haas,
  Kidder, Pfeiffer, Scheel, \& Szilagyi}]{foucart:14a}
Foucart, F., Deaton, M.~B., Duez, M.~D., {et~al.} 2014, \prd, 90, 024026,
  \href{http://arxiv.org/abs/1405.1121}{arXiv:astro-ph.HE/1405.1121}

\bibitem[{{Foucart} {et~al.}(2015{\natexlab{a}}){Foucart}, {Haas}, {Duez},
  {O'Connor}, {Ott}, {Roberts}, {Kidder}, {Lippuner}, {Pfeiffer}, \&
  {Scheel}}]{foucart:15b}
{Foucart}, F., {Haas}, R., {Duez}, M.~D., {et~al.} 2015{\natexlab{a}}, ArXiv
  e-prints,
  \href{http://arxiv.org/abs/1510.06398}{arXiv:astro-ph.HE/1510.06398}

\bibitem[{{Foucart} {et~al.}(2015{\natexlab{b}}){Foucart}, {O'Connor},
  {Roberts}, {Duez}, {Haas}, {Kidder}, {Ott}, {Pfeiffer}, {Scheel}, \&
  {Szilagyi}}]{foucart:15}
{Foucart}, F., {O'Connor}, E., {Roberts}, L., {et~al.} 2015{\natexlab{b}},
  \prd, 91, 124021,
  \href{http://arxiv.org/abs/1502.04146}{arXiv:astro-ph.HE/1502.04146}


\bibitem[{{Freiburghaus} {et~al.}(1999){Freiburghaus}, {Rosswog}, \&
  {Thielemann}}]{freiburghaus:99}
{Freiburghaus}, C., {Rosswog}, S., \& {Thielemann}, F.-K. 1999, \apjl, 525,
  L121, \href{http://doi.org/10.1086/312343}{doi:10.1086/312343}


\bibitem[{{Fuller} \& {Meyer}(1995)}]{fuller:95}
{Fuller}, G.~M., \& {Meyer}, B.~S. 1995, \apj, 453, 792,
  \href{http://doi.org/10.1086/176442}{doi:10.1086/176442}

\bibitem[{Gaburov {et~al.}(2010)Gaburov, Lombardi, \& Zwart}]{Gaburov:2009kg}
Gaburov, E., Lombardi, J., \& Zwart, S.~P. 2010, \mnras, 402, 105,
  \href{http://arxiv.org/abs/0904.0997}{arXiv:astro-ph.SR/0904.0997}

\bibitem[Goriely et al.(2009)]{goriely:09} Goriely, S., Chamel, N., \&
Pearson, J.~M.\ 2009, Physical Review Letters, 102, 152503 

\bibitem[{{Goriely} {et~al.}(2015){Goriely}, {Bauswein}, {Just}, {Pllumbi}, \&
  {Janka}}]{goriely:15}
{Goriely}, S., {Bauswein}, A., {Just}, O., {Pllumbi}, E., \& {Janka}, H.-T.
  2015, \mnras, 452, 3894,
  \href{http://arxiv.org/abs/1504.04377}{arXiv:astro-ph.SR/1504.04377}

\bibitem[{{Hansen} {et~al.}(2014){Hansen}, {Montes}, \& {Arcones}}]{hansen:14}
{Hansen}, C.~J., {Montes}, F., \& {Arcones}, A. 2014, \apj, 797, 123,
  \href{http://arxiv.org/abs/1408.4135}{arXiv:astro-ph.SR/1408.4135}

\bibitem[{{Hebeler} {et~al.}(2013){Hebeler}, {Lattimer}, {Pethick}, \&
  {Schwenk}}]{hebeler:13}
{Hebeler}, K., {Lattimer}, J.~M., {Pethick}, C.~J., \& {Schwenk}, A. 2013,
  \apj, 773, 11,
  \href{http://arxiv.org/abs/1303.4662}{arXiv:astro-ph.SR/1303.4662}

\bibitem[{{Hoffman} {et~al.}(1997){Hoffman}, {Woosley}, \& {Qian}}]{hoffman:97}
{Hoffman}, R.~D., {Woosley}, S.~E., \& {Qian}, Y.-Z. 1997, \apj, 482, 951,
  \href{http://arxiv.org/abs/astro-ph/9611097}{arXiv:astro-ph/9611097}

\bibitem[{{Horowitz}(2002)}]{horowitz:02}
{Horowitz}, C.~J. 2002, \prd, 65, 043001,
  \href{http://arxiv.org/abs/arXiv:astro-ph/0109209}{arXiv:astro-ph/0109209}

\bibitem[{{Hotokezaka} {et~al.}(2013){Hotokezaka}, {Kyutoku}, {Tanaka},
  {Kiuchi}, {Sekiguchi}, {Shibata}, \& {Wanajo}}]{hotokezaka:13b}
{Hotokezaka}, K., {Kyutoku}, K., {Tanaka}, M., {et~al.} 2013, \apjl, 778, L16,
  \href{http://arxiv.org/abs/1310.1623}{arXiv:astro-ph.HE/1310.1623}

\bibitem[{{H{\"u}depohl} {et~al.}(2010){H{\"u}depohl}, {M{\"u}ller}, {Janka},
  {Marek}, \& {Raffelt}}]{huedepohl:10}
{H{\"u}depohl}, L., {M{\"u}ller}, B., {Janka}, H.-T., {Marek}, A., \&
  {Raffelt}, G.~G. 2010, Phys. Rev. Lett., 104, 251101

\bibitem[{{Ishimaru} {et~al.}(2015){Ishimaru}, {Wanajo}, \&
  {Prantzos}}]{ishimaru:15}
{Ishimaru}, Y., {Wanajo}, S., \& {Prantzos}, N. 2015, \apjl, 804, L35,
  \href{http://arxiv.org/abs/1504.04559}{arXiv:1504.04559}

\bibitem[{{Jin} {et~al.}(2015){Jin}, {Li}, {Cano}, {Covino}, {Fan}, \&
  {Wei}}]{jin:15}
{Jin}, Z.-P., {Li}, X., {Cano}, Z., {et~al.} 2015, \apjl, 811, L22,
  \href{http://arxiv.org/abs/1507.07206}{arXiv:astro-ph.HE/1507.07206}

\bibitem[{{Just} {et~al.}(2015){Just}, {Bauswein}, {Pulpillo}, {Goriely}, \&
  {Janka}}]{just:15}
{Just}, O., {Bauswein}, A., {Pulpillo}, R.~A., {Goriely}, S., \& {Janka}, H.-T.
  2015, \mnras, 448, 541,
  \href{http://arxiv.org/abs/1406.2687}{arXiv:astro-ph.SR/1406.2687}

\bibitem[{{Kasen} {et~al.}(2015){Kasen}, {Fern{\'a}ndez}, \&
  {Metzger}}]{kasen:15}
{Kasen}, D., {Fern{\'a}ndez}, R., \& {Metzger}, B.~D. 2015, \mnras, 450, 1777,
  \href{http://arxiv.org/abs/1411.3726}{arXiv:astro-ph.HE/1411.3726}

\bibitem[{{Kiziltan} {et~al.}(2013){Kiziltan}, {Kottas}, {De Yoreo}, \&
  {Thorsett}}]{kiziltan:13}
{Kiziltan}, B., {Kottas}, A., {De Yoreo}, M., \& {Thorsett}, S.~E. 2013, \apj,
  778, 66, \href{http://arxiv.org/abs/1011.4291}{arXiv:1011.4291}

\bibitem[Kolbe et al.(2004)]{kolbe:04} Kolbe, E., Langanke, K., \&
Fuller, G.~M.\ 2004, Physical Review Letters, 92, 111101 

\bibitem[{{Korobkin} {et~al.}(2012){Korobkin}, {Rosswog}, {Arcones}, \&
  {Winteler}}]{korobkin:12}
{Korobkin}, O., {Rosswog}, S., {Arcones}, A., \& {Winteler}, C. 2012, \mnras,
  426, 1940, \href{http://arxiv.org/abs/1206.2379}{arXiv:astro-ph.SR/1206.2379}

\bibitem[Kratz et al.(1993)]{kratz:93} Kratz, K.-L., Bitouzet, J.-P.,
Thielemann, F.-K., Moeller, P., \& Pfeiffer, B.\ 1993, \apj, 403, 216

\bibitem[{{Kyutoku} {et~al.}(2015){Kyutoku}, {Ioka}, {Okawa}, {Shibata}, \&
  {Taniguchi}}]{kyutoku:15}
{Kyutoku}, K., {Ioka}, K., {Okawa}, H., {Shibata}, M., \& {Taniguchi}, K. 2015,
  \prd, 92, 044028,
  \href{http://arxiv.org/abs/1502.05402}{arXiv:astro-ph.HE/1502.05402}


\bibitem[{{Lattimer} {et~al.}(1977){Lattimer}, {Mackie}, {Ravenhall}, \&
  {Schramm}}]{lattimer:77}
{Lattimer}, J.~M., {Mackie}, F., {Ravenhall}, D.~G., \& {Schramm}, D.~N. 1977,
  \apj, 213, 225, \href{http://doi.org/10.1086/155148}{doi:10.1086/155148}

\bibitem[{{Lattimer} \& {Schramm}(1976)}]{lattimer:76}
{Lattimer}, J.~M., \& {Schramm}, D.~N. 1976, \apj, 210, 549,
  \href{http://doi.org/10.1086/154860}{doi:10.1086/154860}

\bibitem[{Lattimer \& Swesty(1991)}]{lseos:91}
Lattimer, J.~M., \& Swesty, F.~D. 1991, {Nucl. Phys. A}, 535, 331

\bibitem[{{Lee} \& {Ramirez-Ruiz}(2007)}]{lee_rev:07}
{Lee}, W.~H., \& {Ramirez-Ruiz}, E. 2007, New Journal of Physics, 9, 17,
  \href{http://arxiv.org/abs/astro-ph/0701874}{arXiv:astro-ph/0701874}

\bibitem[{{Lippuner} \& {Roberts}(2015)}]{lippuner:15}
{Lippuner}, J., \& {Roberts}, L.~F. 2015, \apj, 815, 82,
  \href{http://arxiv.org/abs/1508.03133}{arXiv:astro-ph.HE/1508.03133}

\bibitem[{Lombardi {et~al.}(2006)Lombardi, Proulx, Dooley, Theriault, Ivanova,
  \& Rasio}]{Lombardi:2005nm}
Lombardi, Jr., J.~C., Proulx, Z.~F., Dooley, K.~L., {et~al.} 2006, Astrophys.
  J., 640, 441,
  \href{http://arxiv.org/abs/astro-ph/0509511}{arXiv:astro-ph/astro-ph/0509511}


\bibitem[{{Martin} {et~al.}(2015){Martin}, {Perego}, {Arcones}, {Thielemann},
  {Korobkin}, \& {Rosswog}}]{martin:15}
{Martin}, D., {Perego}, A., {Arcones}, A., {et~al.} 2015, \apj, 813, 2,
  \href{http://arxiv.org/abs/1506.05048}{arXiv:astro-ph.SR/1506.05048}

\bibitem[{{Matteucci} {et~al.}(2014){Matteucci}, {Romano}, {Arcones},
  {Korobkin}, \& {Rosswog}}]{matteucci:14}
{Matteucci}, F., {Romano}, D., {Arcones}, A., {Korobkin}, O., \& {Rosswog}, S.
  2014, \mnras, 438, 2177,
  \href{http://arxiv.org/abs/1311.6980}{arXiv:1311.6980}

\bibitem[Mendoza-Temis et al.(2015)]{mendoza-temis:15} Mendoza-Temis,
J.~d.~J., Wu, M.-R., Langanke, K., et al.\ 2015, \prc, 92, 055805 

\bibitem[{{Metzger} \& {Berger}(2012)}]{metzger:12}
{Metzger}, B.~D., \& {Berger}, E. 2012, \apj, 746, 48,
  \href{http://arxiv.org/abs/1108.6056}{arXiv:astro-ph.HE/1108.6056}

\bibitem[Meyer et al.(1998)]{meyer:98} Meyer, B.~S., McLaughlin,
G.~C., \& Fuller, G.~M.\ 1998, \prc, 58, 3696 
  \href{http://arxiv.org/abs/astro-ph/9809242}{arXiv:astro-ph/9809242}

\bibitem[{{M{\"o}ller} {et~al.}(1997){M{\"o}ller}, {Nix}, \&
  {Kratz}}]{moeller:97}
{M{\"o}ller}, P., {Nix}, J.~R., \& {Kratz}, K.-L. 1997, Atomic Data and Nuclear
  Data Tables, 66, 131,
  \href{http://doi.org/10.1006/adnd.1997.0746}{doi:10.1006/adnd.1997.0746}

\bibitem[{Monaghan(2002)}]{Monaghan:2002ru}
Monaghan, J.~J. 2002, \mnras, 335, 843,
  \href{http://arxiv.org/abs/astro-ph/0204118}{arXiv:astro-ph/astro-ph/0204118}

\bibitem[{{Montes} {et~al.}(2007){Montes}, {Beers}, {Cowan}, {Elliot},
  {Farouqi}, {Gallino}, {Heil}, {Kratz}, {Pfeiffer}, {Pignatari}, \&
  {Schatz}}]{montes:07}
{Montes}, F., {Beers}, T.~C., {Cowan}, J., {et~al.} 2007, \apj, 671, 1685,
  \href{http://arxiv.org/abs/0709.0417}{arXiv:0709.0417}

\bibitem[{{Mumpower} {et~al.}(2015){Mumpower}, {Surman}, {Fang}, {Beard},
  {M{\"o}ller}, {Kawano}, \& {Aprahamian}}]{mumpower:15}
{Mumpower}, M.~R., {Surman}, R., {Fang}, D.-L., {et~al.} 2015, \prc, 92,
  035807, \href{http://arxiv.org/abs/1505.07789}{arXiv:nucl-th/1505.07789}

\bibitem[{Neilsen {et~al.}(2014)Neilsen, Liebling, Anderson, Lehner, O'Connor,
  \& Palenzuela}]{neilsen:14}
Neilsen, D., Liebling, S.~L., Anderson, M., {et~al.} 2014, \prd, 89, 104029,
  \href{http://arxiv.org/abs/1403.3680}{arXiv:1403.3680}

\bibitem[{{O'Connor} \& {Ott}(2010)}]{oconnor:10}
{O'Connor}, E., \& {Ott}, C.~D. 2010, \cqg, 27, 114103

\bibitem[{{{\"O}zel} {et~al.}(2010){{\"O}zel}, {Psaltis}, {Narayan}, \&
  {McClintock}}]{ozel:10bh}
{{\"O}zel}, F., {Psaltis}, D., {Narayan}, R., \& {McClintock}, J.~E. 2010,
  \apj, 725, 1918, \href{http://arxiv.org/abs/1006.2834}{arXiv:1006.2834}

\bibitem[{Paczynski \& Wiita(1980)}]{Paczynski:1979rz}
Paczynski, B., \& Wiita, P.~J. 1980, Astron. Astrophys., 88, 23

\bibitem[{Palenzuela {et~al.}(2015)Palenzuela, Liebling, Neilsen, Lehner,
  Caballero, O’Connor, \& Anderson}]{palenzuela:15}
Palenzuela, C., Liebling, S.~L., Neilsen, D., {et~al.} 2015, Phys. Rev., D92,
  044045, \href{http://arxiv.org/abs/1505.01607}{arXiv:gr-qc/1505.01607}


\bibitem[{Ponce {et~al.}(2012)Ponce, Faber, \& Lombardi}]{Ponce:2011kv}
Ponce, M., Faber, J.~A., \& Lombardi, Jr, J.~C. 2012, \apj, 745, 71,
  \href{http://arxiv.org/abs/1107.1711}{arXiv:astro-ph.CO/1107.1711}

\bibitem[{{Qian}(2000)}]{qian:00}
{Qian}, Y.-Z. 2000, \apjl, 534, L67,
  \href{http://arxiv.org/abs/astro-ph/0003242}{arXiv:astro-ph/0003242}

\bibitem[Qian(2002)]{qian:02} Qian, Y.-Z.\ 2002, \apjl, 569, L103 

\bibitem[{{Qian} \& {Wasserburg}(2008)}]{qian:08}
{Qian}, Y.-Z., \& {Wasserburg}, G.~J. 2008, \apj, 687, 272,
  \href{http://arxiv.org/abs/0807.0809}{arXiv:0807.0809}

\bibitem[{{Radice} {et~al.}(2016){Radice}, {Galeazzi}, {Lippuner}, {Roberts},
  {Ott}, \& {Rezzolla}}]{radice:16b}
{Radice}, D., {Galeazzi}, F., {Lippuner}, J., {et~al.} 2016, ArXiv e-prints,
  \href{http://arxiv.org/abs/1601.02426}{arXiv:astro-ph.HE/1601.02426}

\bibitem[{{Ramirez-Ruiz} {et~al.}(2015){Ramirez-Ruiz}, {Trenti}, {MacLeod},
  {Roberts}, {Lee}, \& {Saladino-Rosas}}]{ramirez-ruiz:15}
{Ramirez-Ruiz}, E., {Trenti}, M., {MacLeod}, M., {et~al.} 2015, \apjl, 802,
  L22, \href{http://arxiv.org/abs/1410.3467}{arXiv:1410.3467}

\bibitem[{{Roberts} {et~al.}(2011){Roberts}, {Kasen}, {Lee}, \&
  {Ramirez-Ruiz}}]{roberts:11}
{Roberts}, L.~F., {Kasen}, D., {Lee}, W.~H., \& {Ramirez-Ruiz}, E. 2011, \apjl,
  736, L21, \href{http://arxiv.org/abs/1104.5504}{arXiv:astro-ph.HE/1104.5504}


\bibitem[{Roederer {et~al.}(2012)Roederer, Lawler, Sobeck, Beers, Cowan,
  Frebel, Ivans, Schatz, Sneden, \& Thompson}]{roederer:12}
Roederer, I.~U., Lawler, J.~E., Sobeck, J.~S., {et~al.} 2012, \apjs, 203, 27

\bibitem[{{Roederer} {et~al.}(2014){Roederer}, {Schatz}, {Lawler}, {Beers},
  {Cowan}, {Frebel}, {Ivans}, {Sneden}, \& {Sobeck}}]{roederer:14}
{Roederer}, I.~U., {Schatz}, H., {Lawler}, J.~E., {et~al.} 2014, \apj, 791, 32,
  \href{http://arxiv.org/abs/1406.4554}{arXiv:astro-ph.SR/1406.4554}

\bibitem[{{Sekiguchi} {et~al.}(2015){Sekiguchi}, {Kiuchi}, {Kyutoku}, \&
  {Shibata}}]{sekiguchi:15}
{Sekiguchi}, Y., {Kiuchi}, K., {Kyutoku}, K., \& {Shibata}, M. 2015, \prd, 91,
  064059, \href{http://arxiv.org/abs/1502.06660}{arXiv:astro-ph.HE/1502.06660}

\bibitem[{{Shen} {et~al.}(2015){Shen}, {Cooke}, {Ramirez-Ruiz}, {Madau},
  {Mayer}, \& {Guedes}}]{shen:15}
{Shen}, S., {Cooke}, R.~J., {Ramirez-Ruiz}, E., {et~al.} 2015, \apj, 807, 115,
  \href{http://arxiv.org/abs/1407.3796}{arXiv:1407.3796}

\bibitem[{{Simmerer} {et~al.}(2004){Simmerer}, {Sneden}, {Cowan}, {Collier},
  {Woolf}, \& {Lawler}}]{simmerer:04}
{Simmerer}, J., {Sneden}, C., {Cowan}, J.~J., {et~al.} 2004, \apj, 617, 1091,
  \href{http://arxiv.org/abs/astro-ph/0410396}{arXiv:astro-ph/0410396}

\bibitem[{Springel \& Hernquist(2002)}]{Springel:2001qb}
Springel, V., \& Hernquist, L. 2002, \mnras, 333, 649,
  \href{http://arxiv.org/abs/0111016}{arXiv:astro-ph/0111016}

\bibitem[{{SXS Collaboration}(2000)}]{SpEC}
{SXS Collaboration}. 2000, Spectral Einstein Code,
  \url{http://www.black-holes.org/SpEC.html}

\bibitem[{{Tanvir} {et~al.}(2013){Tanvir}, {Levan}, {Fruchter}, {Hjorth},
  {Hounsell}, {Wiersema}, \& {Tunnicliffe}}]{tanvir:13}
{Tanvir}, N.~R., {Levan}, A.~J., {Fruchter}, A.~S., {et~al.} 2013, \nat, 500,
  547, \href{http://arxiv.org/abs/1306.4971}{arXiv:astro-ph.HE/1306.4971}

\bibitem[{{The LIGO Scientific Collaboration}(2010)}]{ligo:10}
{The LIGO Scientific Collaboration}. 2010, Classical and Quantum Gravity, 27,
  173001, \href{http://arxiv.org/abs/1003.2480}{arXiv:astro-ph.HE/1003.2480}

\bibitem[{{The LIGO Scientific Collaboration}(2015)}]{aligo}
---. 2015, \cqg, 32, 074001,
  \href{http://arxiv.org/abs/1411.4547}{arXiv:gr-qc/1411.4547}

\bibitem[{{Timmes} \& {Arnett}(1999)}]{timmes:99}
{Timmes}, F.~X., \& {Arnett}, D. 1999, \apjs, 125, 277

\bibitem[{{Travaglio} {et~al.}(2004){Travaglio}, {Gallino}, {Arnone}, {Cowan},
  {Jordan}, \& {Sneden}}]{travaglio:04}
{Travaglio}, C., {Gallino}, R., {Arnone}, E., {et~al.} 2004, \apj, 601, 864,
  \href{http://arxiv.org/abs/astro-ph/0310189}{arXiv:astro-ph/0310189}

\bibitem[{{van de Voort} {et~al.}(2015){van de Voort}, {Quataert}, {Hopkins},
  {Kere{\v s}}, \& {Faucher-Gigu{\`e}re}}]{vandevoort:15}
{van de Voort}, F., {Quataert}, E., {Hopkins}, P.~F., {Kere{\v s}}, D., \&
  {Faucher-Gigu{\`e}re}, C.-A. 2015, \mnras, 447, 140,
  \href{http://arxiv.org/abs/1407.7039}{arXiv:astro-ph.GA/1407.7039}

\bibitem[{{Wanajo} {et~al.}(2014){Wanajo}, {Sekiguchi}, {Nishimura}, {Kiuchi},
  {Kyutoku}, \& {Shibata}}]{wanajo:14}
{Wanajo}, S., {Sekiguchi}, Y., {Nishimura}, N., {et~al.} 2014, \apjl, 789, L39,
  \href{http://arxiv.org/abs/1402.7317}{arXiv:astro-ph.SR/1402.7317}

\bibitem[{{Westin} {et~al.}(2000){Westin}, {Sneden}, {Gustafsson}, \&
  {Cowan}}]{westin:00}
{Westin}, J., {Sneden}, C., {Gustafsson}, B., \& {Cowan}, J.~J. 2000, \apj,
  530, 783,
  \href{http://arxiv.org/abs/astro-ph/9910376}{arXiv:astro-ph/9910376}

\bibitem[{{Woosley} {et~al.}(2002){Woosley}, {Heger}, \& {Weaver}}]{woosley:02}
{Woosley}, S.~E., {Heger}, A., \& {Weaver}, T.~A. 2002, Rev. Mod. Phys., 74,
  1015,
  \href{http://doi.org/10.1103/RevModPhys.74.1015}{doi:10.1103/RevModPhys.74.1015}

\bibitem[{{Yang} {et~al.}(2015){Yang}, {Jin}, {Li}, {Covino}, {Zheng},
  {Hotokezaka}, {Fan}, {Piran}, \& {Wei}}]{yang:15}
{Yang}, B., {Jin}, Z.-P., {Li}, X., {et~al.} 2015, Nature Communications, 6,
  7323, \href{http://arxiv.org/abs/1503.07761}{arXiv:astro-ph.HE/1503.07761}

\end{thebibliography}
\end{document}